\documentclass[reqno,a4paper]{amsart}
\usepackage[active]{srcltx}
\usepackage{fancyhdr}
\usepackage[utf8]{inputenc}
\usepackage{accents}
\DeclareRobustCommand{\SkipTocEntry}[5]{} 
\usepackage{times}
\usepackage{latexsym,amsopn,amssymb,amsmath}
\usepackage{amsthm}
\usepackage{amsfonts,amsbsy,amscd,stmaryrd,bbold}
\usepackage{paralist}
\usepackage{hyperref}
\hypersetup{colorlinks=true,linkcolor=Emerald,urlcolor=blue} 

\usepackage[usenames,dvipsnames]{color}

\usepackage{mathrsfs} 

\newcommand{\N}{\mathbb{N}}

\newcommand{\R}{\mathbb{R}}

\newcommand{\Z}{\mathbb{Z}}

\newcommand{\cb}{\mathcal{B}}

\newcommand{\cd}{\mathcal{D}}
\newcommand{\ce}{\mathcal{E}}
\newcommand{\cf}{\mathcal{F}}
\newcommand{\cg}{\mathcal{G}}
\newcommand{\ch}{\mathcal{H}}

\newcommand{\ck}{\mathcal{K}}

\newcommand{\cs}{\mathcal{S}}

\newcommand{\cu}{\mathcal{U}}

\def\rond{\mathscr}

\newcommand{\rb}{\rond{B}}

\newcommand{\ru}{\rond{U}}

\def\rmc{\mathrm{c}}

\def\d{\mathrm{d}}

\def\rme{\mathrm{e}}
\def\e{\mathrm{e}}

\def\rmi{\mathrm{i}}
\def\i{\mathrm{i}}

\renewcommand{\proof}{\noindent{\bf Proof. }}

\def\braket#1#2{\langle{#1}|{#2}\rangle}

\def\jap#1{\langle {#1} \rangle}

\DeclareMathOperator*{\slim}{s-lim}

\DeclareMathOperator*{\w*lim}{w*-lim}

\def\supp{\mbox{\rm supp\! }}

\def\qed{\hfill $\Box$\medskip}
\def\Div{\mathrm{div}\,}
\long\def\symbolfootnote[#1]#2{\begingroup%
\def\thefootnote{\fnsymbol{footnote}}\footnote[#1]{#2}\endgroup} 

\newtheorem{theorem}{Theorem}[section]
\newtheorem{lemma}[theorem]{Lemma}
\newtheorem{proposition}[theorem]{Proposition}
\newtheorem{corollary}[theorem]{Corollary}

\newtheorem{definition}[theorem]{Definition}
\newtheorem{remark}[theorem]{Remark}

\numberwithin{equation}{section}

\newcommand\ie{{i.\kern2pt e.\ }}

\begin{document}

\title[Limiting Absorption Principle]{On the Limiting absorption
  principle for a new class of Schr\"odinger Hamiltonians
  \\[2mm]
  {\tiny \today}
}

\author[A. Martin]{Alexandre Martin} \address{A. Martin,
  D\'epartement de Math\'ematiques, Universit\'e de Cergy-Pontoise,
  95000 Cergy-Pontoise, France}
\email{alexandre.martin@u-cergy.fr }

\begin{abstract}
  We prove the limiting absorption principle and discuss the
  continuity properties of the boundary values of the resolvent for a
  class of form bounded perturbations of the Euclidean Laplacian
  $\Delta$ that covers both short and long range potentials with an
  essentially optimal behaviour at infinity. For this, we give an extension of Nakamura's results (see \cite{Nak}).
%
\end{abstract}

\maketitle
\tableofcontents 

\section{Introduction}\label{s:intro}
The purpose of this article is to prove a limiting absorption principle for a certain class of Schr\"odinger operator with real potential and to study their essential spectrum. Because this operators are self-adjoint, we already know that their spectrum is in the real axis. We also know that the non negative Laplacian operator $\Delta$ (Schr\"odinger operator with no potential) has for spectrum the real set $[0,+\infty)$ with purely absolutely continuous spectrum on this set. If we add to $\Delta$ a "small" potential (with compact properties with respect to $\Delta$), the essential spectrum of this new operator is the same that $\Delta$ essential spectrum which is continuous. We are interested in the nature of the essential spectrum of the perturbed operator and in the behaviour of the resolvent operator near the essential spectrum.

We will say that a self-adjoint operator has \emph{normal spectrum} in an open real set $O$ if it has no singular continuous spectrum in $O$ and its eigenvalues in $O$ are of finite multiplicities and have no accumulation points inside $O$. Note that if we have a Limiting Absorption Principle for an operator $H$ on $O$, $H$ has normal spectrum on $O$.

A general technique for proving this property is due to E.\
Mourre \cite{Mo1} and it involves a local version of the positive
commutator method due to C.R.~Putnam \cite{Pu1,Pu2}. For various
extensions and applications of these techniques we refer to
\cite{ABG}. Roughly speaking, the idea is to search for a second
self-adjoint operator $A$ such that $H$ is regular in a certain sense
with respect to $A$ and such that 
$H$ satisfies the Mourre estimate on a set $I$ in the following sense
\[E(I)[H,iA]E(I)\geq c_0E(I)+K\]
where $E(I)$ is the spectral measure of $H$ on $I$, $c_0>0$ and $K$ a compact operator. Then one says that the
operator $A$ is \emph{conjugate} to $H$ on $I$. 

When $H=\Delta+V$ is a Schr\"odinger operator, we usually apply the Mourre theorem  with the generator of dilations 
\[A_D=\frac{1}{2}(p\cdot q+q\cdot p),\]
 where
$p=-\rmi\nabla$ and $q$ is the vector of multiplication by
$x$ (see \cite[Proposition 7.4.6]{ABG} and \cite[Section 4]{CFKS}). But in the commutator expressions, derivatives of $V$ appears which can be a problem, if, for example, $V$ has high oscillations at infinity.

In a recent paper S.\ Nakamura \cite{Nak} pointed out the interesting
fact that a different choice of conjugate operator for $H$ can be used to have a limiting absorption principle. This
allows us to avoid imposing conditions on the derivative of the long
range part of the potential. More precisely, if the operator of multiplication by $V(q)$ is $\Delta$-compact
and two other multiplication operators, which include differencies on $V$ and not derivatives, are $\Delta$-bounded, then $H$ has normal
spectrum in $(0,\pi^2/a^2)$ and the limiting absorption principle holds for $H$ locally on this set, outside the eigenvalues. This fact is a consequence of the Mourre
theorem with $A_N$ (see \eqref{eq:def AN}) as conjugate operator. 

Our purpose in this article is to put the results of Nakamura in a more
general abstract setting and get a generalisation of his
result. Moreover, we will show that this generalisation can be applied to potentials for which the Mourre theorem with the generator of dilations as conjugate operator cannot apply (our potentials are not of 
long range type). Furthermore, Nakamura's result cannot apply to this type of potentials which are not $\Delta$-bounded. 
Finally, as usual, we will derive from the limiting absorption principle an application of this theory to wave operators.

We denote $X=\R^\nu$ and $\ch=L^2(X)$. Let $\ch^1$ be the first order
Sobolev space on $X$, denote $\ch^{-1}$ its adjoint space and similarly, we denote $\ch^2$ the second order
Sobolev space on $X$ and $\ch^{-2}$ its adjoint space. All this spaces realised the following
 \[\ch^2\subset\ch^1\subset\ch\subset\ch^{-1}\subset\ch^{-2}.\]Set
$\rb_1=B(\ch^1 ,\ch^{-1})$ and $\rb_2=B(\ch^2 ,\ch^{-2})$.  If needed for clarity, if $u$ is a
measurable function on $X$ we denote $u(q)$ the operator of
multiplication by $u$ whose domain and range should be obvious from
the context. If $a\in X$ let $T_a$ be the operator of translation by
$a$, more precisely $(T_af)(x)=f(x+a)$.

We say that $V\in\rb$, $\rb=\rb_1$ or $\rb=\rb_2$, is a \emph{multiplication operator} if
$V\theta(q)=\theta(q)V$ for any $\theta\in C_\rmc^\infty(X)$.  Note
that $V$ is not necessarily the operator of multiplication by a
function, it could be the operator of multiplication by a distribution
of strictly positive order. For example, in the one dimensional case
$V$ could be equal to the derivative of a bounded measurable
function. 
Anyway, if $V$ is a multiplication operator then there is a uniquely
defined temperate distribution $v$ on $X$ such that $Vf=vf$ for all
$f\in C_\rmc^\infty(X)$ and then $T_a VT_a^*=v(\cdot+a)$. In general
we simplify notations and do not distinguish between the operator $V$
and the distribution $v$, so we write $V=V(q)$ and
$T_a VT_a^*=V(q+a)$. 

We will extend Nakamura's result in two directions. First, we will use the Mourre theory with a class of potential $V:\ch^2\rightarrow L^2$ or $V:\ch^1\rightarrow \ch^{-1}$ (cf. \cite{ABG}) and satisfying a weaker regularity. In particular, this includes potentials with Coulomb singularities, and also short range potentials (see Definition \ref{df:SR}). Secondly, we will use the Mourre theory with a more general class of conjugate operators including $A_N$ (see \eqref{eq:A form}).

Let $a>0$ and let $\sin(ap)=\bigl(\sin(-ia\partial_{x_1}),\cdots,\sin(-ia\partial_{x_\nu})\bigr)$.
 Let
 \begin{equation}\label{eq:def AN}
 A_N=\frac{1}{2}\bigl(\sin(ap)\cdot q+q\cdot \sin(ap)\bigr). 
 \end{equation}
 
Fix a real function $\xi\in C^\infty(X)$ such that $\xi(x)=0$ if
$|x|<1$ and $\xi(x)=1$ if $|x|>2$.

\begin{theorem}\label{th:LAP-intro}
  Let $a\in \R$ and $V:\ch^2\to L^2$ (respectively $V:\ch^1\to\ch^{-1}$) be a compact symmetric multiplication 
  operator with, for all vector $e$ of the canonical basis of  $\R^\nu$,
\begin{equation}\label{eq:LR-intro}
  \int_1^\infty \|\xi(q/r)|q| (V(q+a e) -V(q))\|_{\rb} 
  \frac{\d r}{r} <\infty 
\end{equation}
where $\rb=\rb_2$ (respectively $\rb=\rb_1$).
Then the self-adjoint operator $H=\Delta+V$ on $\ch$ has normal
spectrum in $(0,(\pi/a)^2)$ and, for some appropriate Besov space $\ck$, the limits
\begin{equation}\label{eq:w-lim res}
(H-\lambda\pm\i0)^{-1}:=\w*lim_{\mu\downarrow0}(H-\lambda\pm\rmi\mu)^{-1}
\end{equation}
exist in $\cb(\ck,\ck^*)$, locally uniformly in
$\lambda\in(0,(\pi/a)^2)$ outside the eigenvalues of $H$.
\end{theorem}

We make some comments in connection with the Theorem
\ref{th:LAP-intro}.
\begin{enumerate} 
\item The Besov space is defined in Section \ref{s:not} by \eqref{eq:def K}.

\smallskip

\item Condition \eqref{eq:LR-intro} is satisfied if
  $\|\jap{q}^{\mu}(V(q+a) -V(q))\|_\rb<\infty$ for a fixed \\
  $\mu>1$. To satisfy this conditions, it suffices that
  $\|\jap{q}^{\mu}V\|_\rb<\infty$. In particular, in dimension $\nu\not=2$, if $V$ is a real function on $\R^\nu$ and if there is $\mu>1$ such that  $\bigl(\langle \cdot\rangle^\mu V(\cdot)\bigr)^p$ is in the Kato class, with $p=1$ if
$\nu=1$, and $p=\nu/2$ if $\nu\geq 3$, then condition \eqref{eq:LR-intro} is satisfied (see Proposition \ref{pr:Kato}). If this is the case for all $a\in \R$, then the limiting absorption principle is true on $(0,+\infty)$. 

\smallskip

\item In the case where $V:\ch^2\rightarrow L^2$ is compact, in \cite{Nak}, $V$ is assumed to satisfy $q(V(q+a e)-V(q))$ and $q^2(V(q+a e)+V(q-a e)-2V(q))$ be $\Delta$-bounded. This assumptions implies to the $C^2(A_N,\ch^2,L^2)$ regularity. Observe that, since $q(V(q+a e)-V(q))$ appears in $[V,iA_N]$, \eqref{eq:LR-intro} implies the $C^{1,1}(A_N,\ch^2,\ch^{-2})$ regularity which is implied by the $C^2(A_N,\ch^2,L^2)$ regularity.

\smallskip

\item In one dimension, let $V$ such that 
\begin{equation}\label{eq:exemple1}
\widehat{qV}(\xi)=\sum_{n=-\infty}^{+\infty} \lambda_n \chi(\xi-n),
\end{equation}
where $\widehat{\cdot}$ is the Fourier transform, $\lambda_n\in\R$ and $\chi$ is compactly support, then $V$ satisfies assumptions of Theorem \ref{th:LAP-intro} with $\rb=\rb_1$ but $V$ is neither $\Delta$-bounded nor of class 
$C^1(A_D,\ch^1,\ch^{-1})$. In particular, we can neither apply the Mourre Theorem with the generator of dilation (see \cite[p.258]{ABG}) nor Nakamura's Theorem (see lemma \ref{l:Nak regularity}).

\end{enumerate}

 As in \cite{Nak}, the limiting absorption principle is limited to $(0,(\pi/a)^2)$. The bound $(\pi/a)^2$ is artificial and appears with the choice of vector field $\sin(ap)$.
 In fact, by a simple computation with the Laplacian $\Delta$ in $L^2(\R^\nu)$, we have
 \begin{eqnarray*}
 [\Delta,iA_N]&=&[\Delta,i\frac{1}{2}\bigl(\sin(ap)\cdot q+q\cdot \sin(ap)\bigr)]\\
 &=&\frac{1}{2}\bigl(\sin(ap)\cdot[\Delta, iq]+[\Delta,iq]\cdot \sin(ap)\bigr)\\
 &=& 2p\cdot sin(ap)
 \end{eqnarray*}
which implies a loss of positivity on $(\pi/a)^2$. This is a drawback except if we can apply \eqref{eq:LR-intro} for all $a>0$.
We will use Nakamura's method in a broader framework that allows the removing of this drawback.

Let us denote $H_0=\Delta=p^2$. Then $H_0$ is a self-adjoint operator
in $\ch$ with domain $\ch^2$ which extends to a linear symmetric
operator $\ch^1\to\ch^{-1}$ for which we keep the notation $H_0$.  Let
\emph{$V:\ch^1\to\ch^{-1}$ be a linear symmetric compact operator}.
Then $H=H_0+V$ is a symmetric operator $\ch^1\to\ch^{-1}$ which
induces a self-adjoint operator in $\ch$ for which we keep the
notation $H$. Let $E_0$ and $E$ be the spectral measures of $H_0$ and
$H$.

Note that for each non real $z$ the resolvent $R(z)=(H-z)^{-1}$ of the
self-adjoint operator $H$ in $\ch$ extends to a continuous operator
$R(z):\ch^{-1}\to\ch^{1}$ which is in fact the inverse of the
bijective operator $H-z:\ch^1\to\ch^{-1}$.

$A_N$ and $A_D$ belongs to a general class of conjugate operator, which appears in \cite[Proposition 4.2.3]{ABG}. This is the class of operator which can be written like
\begin{equation}\label{eq:A form}
A_u=\frac{1}{2}\left(u(p)\cdot q+q\cdot u(p) \right)
\end{equation}
 where $u$ is a $C^\infty$ vector field with all the derivates bounded. We will see that this conjugate operator is self-adjoint on some domain (see Section \ref{s:flow}). Conjugate operators of this form were already used in Mourre's
paper \cite[page 395]{Mo1}. 

Remark that the commutator of such conjugate operator with a function of $p$
is quite explicit: denoting $h'=\nabla h$ then 
\begin{equation}\label{eq:com}
[h(p),\rmi A_u]=[h(p),\rmi u(p)q]=u(p)\cdot h'(p) =(u\cdot h')(p) .
\end{equation}
In particular $[H_0,\rmi A_u]=2p\cdot u(p)$. We denote by the same notation $e^{ \rmi \tau A_u}$ the $C_0$-group in $\ch^{1}$ and in $\ch^{-1}$.

 One says that $A$ is
\emph{strictly conjugate to} $H_0$ on $J$ if there is a real number
$a>0$ such that $E_0(J)[H_0,\rmi A]E_0(J)\geq aE_0(J)$, which in our
case means $2k\cdot u(k)\geq a$ for each $k\in X$ such that $|k|^2\in J$. Taking $A_u$ in this class, we have the following                                           

\begin{theorem}\label{th:756}
  Let $V:\ch^1\to\ch^{-1}$ be a compact
  symmetric operator such that there is $u$ a $C^\infty$ bounded vector field with all derivatives bounded such that $V$ is of class $C^{1,1}(A_u,\ch^1,\ch^{-1})$ in
  the following sense:
\begin{equation}\label{eq:c11}
\int_0^1\|V_\tau+V_{-\tau}-2V\|_{\rb_1} \,\frac{\d\tau}{\tau^2}<\infty, 
\quad\text{where } V_\tau=\rme^{\i\tau A_u}V\rme^{-\i\tau A_u} .
\end{equation}
Let $J$ be an open real set such that
\begin{equation}\label{eq:J}
  \inf\{k\cdot u(k) \mid k\in X, |k|^2\in J\}>0 .
\end{equation}
Then $H$ has normal spectrum in $J$ and the limits
\begin{equation}\label{eq:lap}
R(\lambda\pm\i0):=\w*lim_{\mu\downarrow0}R(\lambda\pm\rmi\mu)
\end{equation}
exist in $B\left(\ch^{-1}_{1/2,1},\ch^{1}_{-1/2,\infty}\right)$, locally uniformly in
$\lambda\in J$ outside the eigenvalues of $H$, where $\ch^{-1}_{1/2,1}$ and $\ch^{1}_{-1/2,\infty}$ are interpolation spaces which are defined on Section \ref{s:not}.
\end{theorem}

We make some remarks about this Theorem:
\begin{enumerate}
\item To check the $C^{1,1}(A_u,\ch^1,\ch^{-1})$ property,  it is useful to have $e^{itA_u}\ch^1\subset \ch^1$. For that, we will make a comment in the Section \ref{s:flow} on the flow generated by the vector field $u$ associated to $A_u$.

\smallskip

\item If $k\cdot u(k)$ is positive for all $k\not=0$, Theorem \ref{th:756} applies with $J=(0,+\infty)$.

\smallskip

\item If $V$ is the divergence of a short range potential (see Definition \ref{df:SR}), then Theorem \ref{th:756} applies. A certain class of this type of potential were already studied in \cite{Co} and \cite{CoG}.

\smallskip

\item Since $R(\lambda\pm\i0)$ exists in $B\left(\ch^{-1}_{1/2,1},\ch^{1}_{-1/2,\infty}\right)$, this operator exists in \\$B\left(C^\infty_c(\R^\nu),D'\right)$, with $D'$ the space of distributions.

\smallskip

\item If $V$ can be seen as a compact operator from $\ch^2$ to $L^2$, Theorem \ref{th:756} is still valid if we replace the assumption "$V$ is of class $C^{1,1}(A_u,\ch^1,\ch^{-1})$" by the weak assumption "$V\in C^{1,1}(A_u,\ch^2,\ch^{-2})$" with the same proof.

\smallskip

\item Consider the $\Delta$-compact operator $V(q)$ where \[V(x)=(1-\kappa(|x|))\frac{sin(|x|^\alpha)}{|x|^\beta},\] with $\kappa\in C^\infty_c(\R,\R)$ with $\kappa(|x|)=1$ if $|x|<1$, $0\leq\kappa\leq1$, $\alpha>0$ and $\beta>0$. Note that this type of potential was already studied in \cite{BaD, DMR, DR1, DR2, JMb, ReT1, ReT2}. In \cite{JMb}, they proved that if $|\alpha-1|+\beta>1$, then $V$ has the good regularity with $A_D$ but, if $|\alpha-1|+\beta<1$, $H\notin C^1(A_D)$. In the latter case, we cannot apply the Mourre theory with the generator of dilation. Here, we prove that, with a certain choice of $u$, $V\in C^{1,1}(A_u,\ch^2,\ch^{-2})$ if $2\alpha+\beta>3$ (see lemma \ref{l:oscillant}). In that case, Theorem \ref{th:756} applies. In particular, in the region $2\alpha+\beta>3$ and $\alpha+\beta\leq2$, we have the limiting absorption principle but $H$ is not of class $C^1(A_D)$. In Section \ref{s:potential}, we will see that  Theorem \ref{th:756} also applies if $\beta\leq0$ under certain condition on $\alpha$.

\smallskip

\item Let $\kappa\in C^\infty_c(\R,\R)$ such that $\kappa=1$ on $[-1,1]$ and $0\leq \kappa\leq 1$.
Let
\[
V(x)=(1-\kappa(|x|))\exp(3|x|/4)\sin(\exp(|x|)).
\]
We can show that, for all $u$ bounded, $V\in C^\infty(A_u,\ch^1,\ch^{-1})$ and Theorem \ref{th:756} applies (see Lemma \ref{l:ex potentiel Cinfty}). Moreover, this implies good regularity properties on the boundary values of the resolvent. Since $V$ is not $\Delta$-bounded, we cannot use the $C^1(A_D,\ch^2,\ch^{-2})$ (see \cite[Theorem 6.3.4]{ABG}). \\
We can also prove that $V\notin C^1(A_D,\ch^1,\ch^{-1})$. In particular, Theorem \ref{th:756} does not apply with $A_u=A_D$.

\smallskip
 
\item If $V:\ch^1\rightarrow\ch^{-1}$ is compact and if there is $\mu>0$ such that $x\mapsto \langle x\rangle^{1+\mu} V(x)$ is in $\ch^{-1}$ ($V$ is assumed to be short range in a quite weak sense), then we can apply Theorem \ref{th:756} with an appropriate $u$ (see lemma \ref{l:H-1}). We will provide in this class an concrete example which cannot be treated with the generator of dilations or Nakamura's result (see lemma \ref{l: example V}).

\end{enumerate}

\smallskip

Now we will see a third result concerning existence of wave operators which are useful in scattering theory (see \cite{RS3}).
\begin{definition}\label{df:SR}
  A linear operator $S\in\rb_1$ is \emph{short range} if it is compact,
  symmetric, and
\begin{equation}\label{eq:SR}
  \int_1^\infty\|\xi(q/r)S\|_{\rb_1} \, \d r<\infty .
\end{equation}
\end{definition}

Remark that \eqref{eq:LR-intro} is satisfied if
\begin{equation}\label{eq:srl}
\int_1^\infty
  \|\xi(q/r)|q|V\|_{\rb} \, \frac{\d r}{r}<\infty
\end{equation}
which is a short range type condition.

Note that we do not require $S$ to be local.  Clearly this condition
requires less decay than the condition \eqref{eq:srl}.  Then we
have:

\begin{theorem}\label{th:scatt}
  Let $H$ be as in Theorem \ref{th:LAP-intro} and let $S$ be a short
  range operator.  Then the self-adjoint operator $K=H+S$ has normal
  spectrum in $(0,\infty)$ and the wave operators
\begin{equation}\label{eq:wave}
\Omega_\pm = \slim_{t\to\pm\infty}\e^{\rmi tK} \e^{-\rmi t{H}} E_H^\rmc
\end{equation}
exist and are complete, where $E_H^\rmc$ is the projection onto the
continuity subspace of $H$. 
\end{theorem}

{\bf We now prove Theorem \ref{th:scatt}.} We have $K=\Delta+V+S$ and
from \cite[7.5.8]{ABG} it follows that $S$, hence $V+S$, is of class
$C^{1,1}(A_N,\ch^1,\ch^{-1})$ for all $a>0$ so that we can use Theorem \ref{th:756} to deduce that
$K$ has normal spectrum in $(0,\infty)$ and that the boundary values
of its resolvent exist as in the case of $H$. For the existence and
completeness of the wave operators we use \cite[Proposition
7.5.6]{ABG} with the following change of notations: $H_0,H,V$ from the
quoted proposition are our $H,K,S$ respectively. It remains only to
check that $S$ satisfies the last condition required on $V$ in that
proposition: but this is a consequence of \cite[Theorem
2.14]{GM}. \qed

We will give on Section \ref{s:lap} more explicit conditions which ensure that the assumptions of Theorem \ref{th:LAP-intro} and \ref{th:scatt} are satisfied in the case where $V$ and $S$ are real functions.

We make two final remarks. First, the assumption of compactness of $V$
and $S$ as operators $\ch^1\to\ch^{-1}$ is too strong for some
applications, for example it is not satisfied if \\
$X=\R^3$ and $V(x)$
has local singularities of order $|x|^{-2}$. But compactness can be
replaced by a notion of smallness at infinity similar to that used in
\cite{GM} which covers such singularities and the arguments there
extend to the present setting. Second, let us mention that we treat
only the case when $H_0$ is the Laplacian $\Delta=p^2$ but an
extension to more general functions $h(p)$ is straightforward with the same class of conjugate operator $A_u$.

\smallskip

The paper is organized as follows. In Section \ref{s:not}, we will give some notations we will use below and we recall some basic fact about regularity with respect to a conjugate operator. In Section \ref{s:gen}, we will prove Theorem \ref{th:756} and extend Nakamura's results by geting properties about the boundary values of the resolvent. In Section \ref{s:lap} , we will give an extension of Nakamura's theorem by using the Mourre theory with $C^{1,1}$ regularity with respect to the conjugate operator $A_N$. In Section \ref{s:potential}, we will give some examples of potentials which satisfies Theorem \ref{th:LAP-intro} and Theorem \ref{th:756} and which are not covered by Mourre Theorem with the generator of dilation and Nakamura's Theorem. In Section \ref{s:flow}, we will study the flow associated to the unitary group generated by $A_u$.

%
%

\section{Notation and basic notions}\label{s:not}

\subsection{Notation}
Let $X=\R^\nu$ and for $s\in\R$ let $\ch^s$ be the usual Sobolev
spaces on $X$ with $\ch^0=\ch=L^2(X)$ whose norm is denoted
$\|\cdot\|$. We are mainly interested in the space $\ch^{1}$ defined
by the norm $\|f\|_1^2=\int\left(|f(x)|^2+|\nabla f(x)|^2\right)\d x$ and
its dual space $\ch^{-1}$.  

Recall that we set $\rb_1=B(\ch^1 ,\ch^{-1})$ and $\rb_2=B(\ch^2 ,\ch^{-2})$ which are Banach spaces 
with norm $\|\cdot\|_\rb$, $\rb=\rb_1,\rb_2$. These spaces satisfy $\rb_1\subset\rb_2$.



We denote $q_j$ the operator of multiplication by the coordinate $x_j$
and $p_j=-\i\partial_j$ considered as operators in $\ch$. For $k\in X$ we denote
$k\cdot q=k_1q_1+\dots+k_\nu q_\nu$. If $u$ is a
measurable function on $X$ let $u(q)$ be the operator of
multiplication by $u$ in $\ch$ and $u(p)=\cf^{-1}u(q)\cf$, where $\cf$
is the Fourier transformation:
\[
(\cf f)(\xi)=(2\pi)^{-\frac{\nu}{2}} \int \rme^{-\rmi x\cdot\xi} u(x) \d x .
\]
If there is no ambiguity we keep the same notation for these operators
when considered as acting in other spaces.  

Throughout this paper $\xi\in C^\infty(X)$ is a real function such
that $\xi(x)=0$ if $|x|<1$ and $\xi(x)=1$ if $|x|>2$. Clearly the
operator $\xi(q)$ acts continuously in all the spaces $\ch^s$.

We are mainly interested in potentials $V$ which are multiplication
operators in the following more general sense.

\begin{definition}\label{df:mult} A map
  $V\in\rb$ is called a \emph{multiplication operator} if
  $V\rme^{\i k\cdot q}=\rme^{\i k\cdot q}V$ for all $k\in X$. Or, equivalently, if
  $V\theta(q)=\theta(q)V$ for all $\theta\in C_\rmc^\infty(X)$.
\end{definition}

For the proof of the equivalence, note first that from
$V\rme^{\i k\cdot q}=\rme^{\i k\cdot q}V, \forall k\in X$ we get
$V\theta(q)=\theta(q)V$ for any Schwartz test function $\theta$
because
$(2\pi)^{\frac{\nu}{2}}\theta(q)= \int \rme^{\rmi kq} (\cf\theta)(k)
\d k$
and second that if $\eta\in C^1(X)$ is bounded with bounded derivative
then $\eta(q)$ is the strong limit in $\rb$ of a sequence of
operators $\theta(q)$ with $\theta\in C_\rmc^\infty(X)$.

As we mentioned in the introduction, such a $V$ is necessarily the
operator of multiplication by a distribution that we also denote $V$
and we sometimes write the associated operator $V(q)$. For example,
the distribution $V$ could be the divergence $\Div W$ of a measurable
vector field $W:X\to X$ such that multiplication by the components of
$W$ sends $\ch^1$ into $\ch$. For example, $W$ could be a bounded
function and if this function tends to zero at infinity then $V$ will
be a compact operator $\ch^1\to\ch^{-1}$. we say that a multiplication operator $V$ is $\Delta$-compact if $V:\ch^2\rightarrow L^2$ is a compact operator.


As usual $\jap{x}=\sqrt{1+|x|^2}$. Then $\jap{q}$ is the operator of
multiplication by the function $x\mapsto\jap{x}$ and
$\jap{p}=\cf^{-1}\jap{q}\cf$.  For real $s,t$ we denote $\ch^t_s$ the
space defined by the norm
\begin{equation}\label{eq:K}
\|f\|_{\ch^t_s}= \|\jap{q}^s f\|_{\ch^t}= \|\jap{p}^t \jap{q}^s f\|=\| \jap{q}^s\jap{p}^t f\| .
\end{equation}
Note that the adjoint space of $\ch^t_s$ may be identified with
$\ch^{-t}_{-s}$.

A finer Besov type version $\ch^{-1}_{1/2,1}$ of $\ch^{-1}_{1/2}$
appears naturally in the theory.  To alleviate the writing we denote
it $\ck$. This space is defined by the norm
\begin{equation}\label{eq:def K}
\|f\|_\ck=\|\theta(q)f\|_{\ch^{-1}}+
\int_1^\infty\|\tau^{1/2}\psi(q/\tau)f\|_{\ch^{-1}} \frac{\d\tau}{\tau}  
\end{equation}
where $\theta,\psi\in C_\rmc^\infty(X)$ with $\theta(x)=1$ if $|x|<1$,
$\psi(x)=0$ if $|x|<1/2$, and $\psi(x)=1$ if $1<|x|<2$.  The adjoint
space $\ck^*$ of $\ck$ is the Besov space $\ch^{1}_{-1/2,\infty}$ (see
\cite[Chapter 4]{ABG}).

We will see in Section \ref{s:flow} that if $u:X\to X$ is a $C^\infty$ vector field all of whose derivatives
are bounded then the operator 
\begin{equation}\label{eq:A}
A_u=\frac{1}{2}\left(u(p)\cdot q+q\cdot u(p) \right)
=\frac{1}{2}\sum_{j=1}^\nu \left(u_j(p)q_j+q_ju_j(p) \right)
=u(p)\cdot q+\frac{\rmi}{2}(\mathrm{div} u)(p)
\end{equation}
with domain $C_{\rmc}^\infty(X)$ is essentially self-adjoint in $\ch$;
we keep the notation $A_u$ for its closure. 
Remark that the unitary group
$\rme^{\rmi\tau A_u}$ generated by $A_u$ leaves invariant all the spaces
$\ch^s_t$ and $\ck$ (see Section \ref{s:flow}).

Since we will use a lot the case of $u$ bounded, let $\cu$ be the space of vector fields $u$ bounded with all derivatives bounded such that $x\cdot u(x)>0$ for all $x\not=0$.

\subsection{Regularity}

Let $F', F''$ be two Banach space and $T:F'\rightarrow F''$ a bounded operator.

Let $A$ a self-adjoint operator.

Let $k\in\N$. we say that $T\in C^k(A,F',F'')$ if, for all $f\in F'$, the map $\R\ni t\rightarrow e^{itA}Te^{-itA}f$ has the usual $C^k$ regularity. The following characterisation is available:

\begin{proposition}
 $T\in C^1(A,F',F'')$ if and only if $[T,A]$ has an extension in $\cb(F',F'')$.
\end{proposition}
It follows that, for $k>1$, $T\in C^k(A,F',F'')$ if and only if $[T,A]\in C^{k-1}(A,F',F'')$.


We can define another class of regularity called the $C^{1,1}$ regularity:

\begin{proposition}
we say that $T\in C^{1,1}(A,F',F'')$ if and only if 
\[
\int_0^1\|T_\tau+T_{-\tau}-2T\|_{\cb(F',F'')} \,\frac{\d\tau}{\tau^2}<\infty, 
\]
 where $T_\tau=\rme^{\i\tau A_u}T\rme^{-\i\tau A_u}$.
\end{proposition} 

An easier result can be used:
\begin{proposition}[Proposition 7.5.7 from \cite{ABG}]
Let $\xi\in C^\infty(X)$ such that $\xi(x)=0$ if
$|x|<1$ and $\xi(x)=1$ if $|x|>2$.
If $T$ satisfies
\[\int_1^\infty \|\xi(q/r)[T,iA]\|_{\cb(F',F'' )}
  \frac{\d r}{r} <\infty\]
  then $T$ is of class $C^{1,1}(A,F',F'')$.
\end{proposition} 

If $T$ is not bounded, we say that $T\in C^k(A,F',F'')$ if for $z\notin\sigma(T)$, $(T-z)^{-1}\in C^k(A,F',F'')$.

\begin{proposition}
For all $k>1$, we have
\[ C^{k}(A,F',F'')\subset C^{1,1}(A,F',F'')\subset C^{1}(A,F',F'').\]
\end{proposition}
If $F'=F''=\ch$ is an Hilbert space, we note $C^1(A)=C^1(A,\ch,\ch^*)$.
If $T$ is self-adjoint, we have the following:

\begin{theorem}[Theorem 6.3.4 from \cite{ABG}]\label{th:634}
Let $A$ and $T$ be self-adjoint operator in a Hilbert space $\ch$. Assume that the unitary group $\{\exp(iA\tau)\}_{\tau\in\R}$ leaves the domain $D(T)$ of $T$ invariant. Set $\cg=D(T)$ endowed with it graph topology. Then 
\begin{enumerate}
\item $T$ is of class $C^1(A)$ if and only if $T\in C^1(A,\cg,\cg^*)$.

\item $T$ is of class $C^{1,1}(A)$ if and only if $T\in C^{1,1}(A,\cg,\cg^*)$.
\end{enumerate}
\end{theorem}
Remark that, if $T:\ch\rightarrow\ch$ is not bounded, since $T:\cg\rightarrow\cg^*$ is bounded, in general, it is easier to prove that $T\in C^1(A,\cg,\cg^*)$ than $T\in C^1(A)$.

If $\cg$ is the form domain of $H$, we have the following:
\begin{proposition}[see p. 258 of \cite{ABG}]
Let $A$ and $T$ be self-adjoint operators in a Hilbert space $\ch$. Assume that the unitary group $\{\exp(iA\tau)\}_{\tau\in\R}$ leaves the form domain $\cg$ of $T$ invariant. Then 
\begin{enumerate}
\item $T$ is of class $C^k(A)$ if and only if $T\in C^k(A,\cg,\cg^*)$, for all $k\in\N$.

\item $T$ is of class $C^{1,1}(A)$ if and only if $T\in C^{1,1}(A,\cg,\cg^*)$.
\end{enumerate}
\end{proposition}

As previously, since $T:\cg\rightarrow\cg^*$ is always bounded, it is, in general, easier to prove that $T\in C^k(A,\cg,\cg^*)$ than $T\in C^k(A)$.
 
\section{Nakamura's ideas in a more general setting: Boundary values of the resolvent} {\label{s:gen}

In this section, we will prove the Theorem \ref{th:756} and we will see how the regularity of the potential, in relation to $A_u$, can implies a good regularity for the boundary values of the resolvent.

\smallskip

{\bf We now prove Theorem \ref{th:756}} By taking into account the equation \eqref{eq:com}
and the statement of Theorem 7.5.6 in \cite{ABG} we only have to
explain how the space $\ck$ introduced before \eqref{eq:def K} appears into the picture
(this is not the space also denoted $\ck$ in \cite{ABG}). In fact the
quoted theorem gives a more precise result, namely instead of our
$\ck$ one may take the real interpolation space
$\big(D(A_u,\ch^{-1}),\ch^{-1}\big)_{1/2,1}$, where $D(A_u,\ch^{-1})$ is
the domain of the closure of $A_u$ in $\ch^{-1}$. From \eqref{eq:A} and
since $u$ is bounded with all its derivatives bounded it follows
immediately that $D(A_u,\ch^{-1})$ contains the domain of $\jap{q}$ in
$\ch^{-1}$, which is $\ch^{-1}_{1}$. Hence
$\big(D(A_u,\ch^{-1}),\ch^{-1}\big)_{1/2,1}$ contains
$\big(\ch^{-1}_{1},\ch^{-1}\big)_{1/2,1}$ which is
$\ch^{-1}_{1/2,1}=\ck$. \qed

We say that $V$ is of class $C^k(A_u,\ch^1,\ch^{-1})$ for some integer $k\geq1$ if the
map $\tau\mapsto V_\tau\in B(\ch^1,\ch^{-1})$ is $k$ times strongly
differentiable.  We clearly have $C^2(A_u,\ch^1,\ch^{-1})\subset C^{1,1}(A_u,\ch^1,\ch^{-1})$. In \cite{Mo1} the Limiting Absorption Principle is proved essentially for $V\in C^2(A)$ (see \cite{GG} for more details); notice
that in \cite{Mo2} the limiting absorption principle is proved in a
space better (i.e\ larger) than $\ck$ (see \eqref{eq:K}), but not of Besov type.

If $s>1/2$ then $\ch^{-1}_s\subset\ck$ with a continuous and dense
embedding. Hence: 

\begin{corollary}\label{co:756s}
  For each $s>1/2$ the limit
  $R(\lambda\pm\i0)=\w*lim_{\mu\downarrow0}R(\lambda\pm\rmi\mu)$ exists
   in the spaces $B\left(\ch^{-1}_s,\ch^1_{-s}\right)$, locally
  uniformly in $\lambda\in J$ outside the eigenvalues of $H$.
\end{corollary}

The $C^{1,1}(A_u)$ regularity condition \eqref{eq:c11} on $V$ is not
explicit enough for some applications. We now give a simpler condition
which ensures that \eqref{eq:c11} is satisfied.

%
%
%
%

We recall some easily proven facts concerning the class $C^1(A_u,\ch^1,\ch^{-1})$.
First, $V$ is of class $C^1(A_u,\ch^1,\ch^{-1})$ if and only if the function
$\tau\mapsto V_\tau\in B(\ch^1,\ch^{-1})$ is (norm or strongly)
Lipschitz. Notice that we used $e^{i\tau A}\ch^1\subset\ch^1$ to prove this. Second, note that for an arbitrary $V$ the expression
$[V,\rmi A_u]$ is well-defined as symmetric sesquilinear form on
$C_\rmc^\infty(X)$ and $V$ is of class $C^1(A_u,\ch^1,\ch^{-1})$ if and only if this
form is continuous for the topology induced by $\ch^1$. In this case
we keep the notation $[V,\rmi A_u]$ for its continuous extension to
$\ch^1$ and for the continuous symmetric operator $\ch^1\to\ch^{-1}$
associated to it. 

\smallskip

As a consequence of Proposition 7.5.7 from \cite{ABG} with the choice
$\Lambda=\jap{q}$ we get:

\begin{proposition}\label{pr:lr}
  Let $V:\ch^1\to\ch^{-1}$ be a symmetric bounded operator of class \\ $C^1(A_u,\ch^1,\ch^{-1})$
  such that 
\begin{equation}\label{eq:lr}
\int_1^\infty\|\xi(q/r)[V,\rmi A_u]\|_{\rb}
\frac{\d r}{r} <\infty 
\end{equation}
with $\rb=\rb_1$.
Then $V$ is of class $C^{1,1}(A_u,\ch^1,\ch^{-1})$.
\end{proposition}


If the potential $V$ is of a higher regularity class with respect to
$A_u$ then, by using results from \cite{BoG}, we also get an optimal
result on the order of continuity of the boundary values of the
resolvent $R(\lambda\pm\i0)$ as functions of $\lambda$. From
\cite{Mo2} and the improvements in \cite{BGS} one may also get a
precise description of the propagation properties of the dynamical
group $\rme^{\i tH}$ in this context, but we shall not give the
details here. 

To state this regularity result we recall the definition of the
\emph{H\"older-Zygmund continuity classes} of order
$s\in\,(0,\infty)\,$.  Let $\ce$ be a Banach space and $F:\R\to\ce$ a
continuous function. If $0<s<1$ then $F$ is of class $\Lambda^s$ if
$F$ is H\"older continuous of order $s$.  If $s=1$ then $F$ is of
class $\Lambda^1$ if it is of Zygmund class, i.e.
$\|F(t+\varepsilon)+F(t-\varepsilon)-2F(t)\|\leq C\varepsilon$ for all
real $t$ and $\varepsilon>0$. If $s>1$, let us write $s=k+\sigma$ with
$k\geq1$ integer and $0<\sigma\leq1$; then $F$ is of class $\Lambda^s$
if $F$ is $k$ times continuously differentiable and $F^{(k)}$ is of
class $\Lambda^\sigma$.  The corresponding local classes are defined
as follows: if $F$ is defined on an open real set $U$ then $F$ is
\emph{locally of class} $\Lambda^s$ if $\theta F$ is of class
$\Lambda^s$ for any $\theta\in C_\rmc^\infty(U)$.

We say that $V$ is of class $\Lambda^s(A_u,\ch^1,\ch^{-1})$ if the function
$\tau\mapsto V_\tau\in \rb_1$ is of class $\Lambda^s$. We mention that
in a more general context this class is denoted by
$C^{s,\infty}(A_u,\ch^1,\ch^{-1})$, but this does not matter here. In any case, one may
easily check that \\$\Lambda^s(A_u,\ch^1,\ch^{-1})\subset C^{1,1}(A_u,\ch^1,\ch^{-1})\subset C^1(A_u,\ch^1,\ch^{-1})$ if
$s>1$. If $s\geq1$ is an integer then $C^s(A_u,\ch^1,\ch^{-1})\subset\Lambda^s(A_u,\ch^1,\ch^{-1})$
strictly.

\begin{theorem}\label{th:reg}
  Assume that $u$ and $J$ are as in Theorem \ref{th:756} and let $s$
  be a real number such that $s>1/2$. If $V:\ch^1\rightarrow\ch^{-1}$ is a compact symmetric operator of class
  $\Lambda^{s+1/2}(A_u,\ch^1,\ch^{-1})$ then the functions
\begin{equation}\label{eq:reg}
\lambda\mapsto R(\lambda\pm\i0)\in B(\ch^{-1}_s,\ch^1_{-s})
\end{equation}
are locally of class $\Lambda^{s-1/2}$ on $J$ outside the eigenvalues
of $H$. 
\end{theorem}

\proof We shall deduce this from the theorem on page 12 of
\cite{BoG}. First, note that $H$ has a spectral gap because $H_0\geq0$
and $(H+\rmi)^{-1}-(H_0+\rmi)^{-1}$ is a compact operator hence $H$
and $H_0$ have the same essential spectrum. Thus we may use the quoted
theorem and we get the assertion of the present theorem but with
$B(\ch^{-1}_s,\ch^1_{-s})$ replaced by $B(\ch_s,\ch_{-s})$. Then it
suffices to observe that if $z$ belongs to the resolvent set of $H$
then we have
\[
{R}(z)={R}(\rmi) + (z-\rmi) {R}(\rmi)^2 
+ (z-\rmi)^2 {R}(\rmi) {R}(z) {R}(\rmi)
\]
and to note that $R(\rmi)$ sends $\ch^{-1}_s$ into $\ch^1_s$.  \qed

We state explicitly the particular case corresponding to the Mourre
condition \\$V\in C^2(A_u,\ch^1,\ch^{-1})$. We mention that this is equivalent to the
fact that the sesquilinear form $[[V,A_u],A_u]$, which is always well-defined on $C_\rmc^\infty(X)$, extends to a continuous sesquilinear
form on $\ch^1$.

\begin{corollary}\label{co:regm}
  Assume that we are in the conditions of the Theorem \ref{th:756} but
  with the condition \eqref{eq:c11} replaced by the stronger one
  $V\in C^2(A_u,\ch^1,\ch^{-1})$. Then the map \eqref{eq:reg} is H\"older continuous of
  order $s-1/2$ for all $s$ such that $1/2<s<3/2$ and if $s=3/2$ then
  the map \eqref{eq:reg} is of Zygmund class (but could be nowhere
  differentiable).
\end{corollary}

Previously, we saw that $H_0$ verified the Mourre estimate on $I$ if and only if $k\cdot u(k)>0$ for all $k$ such that $|k|^2\in I$. Moreover, we saw that if $H_0$ verified the Mourre estimate on $I$, because the potential $V$ is $H_0$-compact or compact on $\ch^1$, the form domain of $H_0$, to $\ch^{-1}$, $H$ verified the Mourre estimate on the same interval $I$.

Because \[u_N(x)=(\sin(ax_j))_{j=1,\cdots,\nu}\] we have the Mourre estimate only on $I_a=(0,(\pi/a)^2)$, i.e. where $u_N(x)\not=0$; this function constructs some artificial thresholds. If we can choose a vector field $u$ such that $x\cdot u(x)>0$ if $x\not=0$ which satisfied some good conditions of regularity for the potential $V$, we can extend the interval $I_a$ to $I=(0,+\infty)$. For example, we can choose  the vector field $u(x)=(\arctan(x_j))_{j=1,\cdots,\nu}$, the function $\arctan$ being non zero for $x\not=0$, or $u(x)=x/\langle x\rangle$. In particular, with this type of vector field, if $\langle p\rangle^{-1}qV\jap{p}^{-1}$ is bounded, then $V\in C^1(A_u,\ch^1,\ch^{-1})$ and we have the Mourre estimate on all compact subset of $(0,+\infty)$.


\section[An extension of Nakamura's results]{An extension of Nakamura's results}
\label{s:lap}

In this section, we will prove Theorem \ref{th:LAP-intro} and we will give some conditions easy to verify which assure that assumptions of Theorem \ref{th:LAP-intro} and \ref{th:scatt} are satisfied. Moreover, we will give a stronger version of Nakamura's Theoerem with estimates on the boundary values of the resolvent.

In all this section, $\rb=\rb_1=B(\ch^1,\ch^{-1})$.

 We fix a real number $a>0$ and denote
$I_a=\,(0,\frac{\pi^2}{a^2})\,$.  We apply the general results from
Section \ref{s:gen} with the vector field $u$ as in \cite{Nak}:
\begin{equation}\label{eq:nak}
u_N(p)=\left(\sin(ap_1), \dots,\sin(ap_\nu)\right) .
\end{equation}
Then 
\(
(\Div u_N)(p)={\textstyle\sum_j}\partial_{p_j}\sin(ap_j)
={\textstyle\sum_j} a \cos(ap_j)
\) 
hence
\begin{equation}\label{eq:anak}
2A_N={\textstyle\sum_j}\big( q_j\sin(ap_j) + \sin(ap_j) q_j\big)=
{\textstyle\sum_j}\big( 2q_j\sin(ap_j)-\rmi a \cos(ap_j) \big) .
\end{equation}
The operator $A_N$ behaves well with respect to the tensor factorization
$L^2(X)=L^2(\R)^{\otimes\nu}$ and this simplifies the
computations. Indeed, if we denote $B$ the operator $A$ acting in
$L^2(\R)$ we have $A_N=A_1+\dots+A_\nu$ with
$A_1=B\otimes1\dots\otimes1$, $A_2=1\otimes B\otimes 1\dots\otimes1$,
etc.

Let $T_j=\rme^{\i a p_j}$ be the operator of translation by $a$ in the
$j$ direction, i.e. \\$(T_jf)(x)=f(x+a e_j)$ where $e_1,\dots,e_\nu$ is
the natural basis of $X=\R^\nu$.  For any $V:\ch^1\to\ch^{-1}$ set
\begin{equation}\label{eq:dj}
\delta_j(V)=T_jVT_j^*-V
\end{equation}
which is also an operator $\ch^1\to\ch^{-1}$ hence we may consider
$\delta_k\delta_j(V)$, etc. If $V=V(q)$ is a multiplication operator
then
\[
\delta_j(V)=V(q+ae_j)-V(q). 
\]
Remark that, when $V$ is a multiplication operator, $\delta_j(V)$ appears in the first commutator \\
$[V,i A_N]$. The operation $\delta_j$ can also be applied to various unbounded
operators, for example we obviously have $\delta_j(q_k)=a\delta_{jk}$,
where $\delta_{jk}$ is the Kronecker symbol, and \\$\delta_j(u(p))=0$.

If $S\in\rb$ then $[q_j,S]$ and $q_jS$ are well-defined as
sesquilinear forms on $C^\infty_\rmc(X)$ and we say that one of these
expressions is a bounded operator $\ch^1\to\ch^{-1}$ if the
corresponding form is continuous in the topology induced by $\ch^1$.

\begin{theorem}\label{th:LAP}
  Let $V:\ch^1\to\ch^{-1}$ be a compact symmetric operator such that
  for any $j$ the forms $[q_j,V]$ and $q_j\delta_j(V)$ 
  are bounded operators $\ch^1\to\ch^{-1}$ and
\begin{equation}\label{eq:LR}
\int_1^\infty \Big(
\|\xi(q/r)[q_j,V]\|_\rb +\|\xi(q/r)q_j\delta_j(V)\|_\rb
\Big) \frac{\d r}{r} <\infty . 
\end{equation}
Then $H$
has normal spectrum in $I_a$
and the limits $R(\lambda\pm\i0)=\w*lim_{\varepsilon\downarrow
  0}R(\lambda\pm\i\varepsilon)$ exist in
$B(\ck,\ck^*)$,
locally uniformly in $\lambda\in
I_a$ outside the set of eigenvalues of $H$.
\end{theorem}

\begin{remark}
This Theorem give a stronger result than Theorem \ref{th:LAP-intro}, since, if $V$ is a multiplication operator, $[q_j,V]=0$ and \eqref{eq:LR} reduces to \eqref{eq:LR-intro}.
\end{remark}

\proof This is a consequence of Theorem \ref{th:756} and Proposition
\ref{pr:lr} once we have checked that $V$
is of class $C^1(A_N)$ and the relation \eqref{eq:lr} is satisfied. In
order to prove that \\$V\in C^1(A_N,\ch^1,\ch^{-1})$ it suffices to show that the
sesquilinear form on $C_\rmc^\infty(X)$ defined by 
\begin{equation}\label{eq:av}
[2A_N,V]={\textstyle\sum_j} \big(2[q_j\sin(ap_j),V]-\rmi
a[\cos(ap_j),V] \big)
\end{equation}
is continuous for the $\ch^1$ topology. This is clear for the second
term in the sum and for the first one we use
\begin{equation}\label{eq:av1}
[q_j\sin(ap_j),V]= [q_j,V]\sin(ap_j) + q_j[\sin(ap_j),V] .
\end{equation}
The first term on the right hand side defines a bounded operator
$\ch^1\to\ch^{-1}$ by one of the hypotheses of the theorem. For the
second one we first note that 
\begin{equation}\label{eq:aux}
[T_j,V]=\delta_j(V)T_j \quad\text{and}\quad  
[T_j^*,V]=-T_j^*\delta_j(V)
\end{equation}
from which we get 
\begin{equation}\label{eq:av2}
2\rmi [\sin(ap_j),V]=[T_j-T_j^*,V]=\delta_j(V)T_j+T_j^*\delta_j(V)
\end{equation}
from which it follows easily that $q_j [\sin(ap_j),V]$ is
bounded. \\
Thus $V$ is of class 
$C^1(A_N,\ch^1,\ch^{-1})$. 

It remains to show that \eqref{eq:lr} is satisfied. We use
\eqref{eq:av} again: the terms with $\sin(ap_j)$ are treated with the
help of \eqref{eq:av1} and \eqref{eq:av2}. The term with $\cos(ap_j)$
is treated similarly by using
\begin{equation}\label{eq:av3}
2 [\cos(ap_j),V]=[T_j+T_j^*,V]=\delta_j(V)T_j-T_j^*\delta_j(V) .
\end{equation}
Using \eqref{eq:LR}, this proves \eqref{eq:lr}. \qed

\begin{remark}\label{re:cont}{\rm The ``usual'' version of the
    preceding theorem involves derivatives $[p_j,V]$ of the
    potential, instead of the finite differences $\delta_j(V)$, cf.
    \cite[Theorem 7.6.8]{ABG}. Note that the quoted theorem is a
    consequence of Theorem \ref{th:LAP} because
    $\|\delta_j(V)\|_\rb\leq a \|[p_j,V]\|_\rb$.  }\end{remark}

The condition \eqref{eq:LR} says that the operators $[q_j,V]$ and
$q_j\delta_j(V)$ are not only bounded as maps $\ch^1\to\ch^{-1}$ but
also tend to zero at infinity in some weak sense. Then it is clear
that the maps $\lambda\mapsto R(\lambda\pm\rmi0)\in B(\ck,\ck^*)$ are
strongly continuous outside the eigenvalues of $H$, but nothing else
can be said in general.  Stronger conditions on this decay improve the
smoothness properties of the boundary values $R(\lambda\pm\rmi0)$ as
maps $\ch^1_s\to\ch^{-1}_{-s}$. This question is solved in general by
using Theorem \ref{th:reg} but here we consider only a particular case
as an example. One may see in \cite[Theorem 1.7]{GM} the type of
assumptions $V$ has to satisfy in order to improve the smoothness
properties of the boundary values.

Remark that if $V$ is a multiplication operator, we have
\[
\delta_j\delta_k(V)=V(q+ae_j+ae_k)-V(q+ae_j)-V(q+ae_k)+V(q)
\]
which appears in the second commutator $[[V,iA_N],iA_N]$.

The next result is an extension of \cite[Theorem 1]{Nak}: we make the
regularity assumption $V\in C^2(A_N)$ but $V$ is not necessarily an
operator in $\ch$ and we give the precise H\"older continuity order of
the boundary values. 

\begin{theorem}\label{th:nak}
  Let $V=V(q):\ch^1\to\ch^{-1}$ be a symmetric compact multiplication
  operator. Assume that there is a real number $a>0$ such that for all
  $j,k$
\begin{equation*}\label{eq:hyp}
\leqno{\mathbf{(H)}}
\hspace{10mm}
\left\{
\begin{array}{ll}
q_j \big(V(q+ae_j)-V(q)\big) \text{ and } \\[1mm]
q_jq_k\big(V(q+ae_j+ae_k)-V(q+ae_k)-V(q+ae_k)+V(q)\big)
\end{array}
\right.
\end{equation*}
are bounded operators $\ch^1\to\ch^{-1}$. Then $H$ has normal spectrum
in the interval $I_a$ and the limits
$R(\lambda\pm\i0)=\w*lim_{\varepsilon\downarrow
  0}R(\lambda\pm\i\varepsilon)$
exist in $B\left(\ck,\ck^*\right)$, locally uniformly in
$\lambda\in I_a$ outside the eigenvalues of $H$.  If
$\frac{1}{2}<s<\frac{3}{2}$ then the operators
$R(\lambda\pm\i0) \in B(\ch^{-1}_s,\ch^{1}_{-s})$ are locally H\"older
continuous functions of order $s-\frac{1}{2}$ of the parameter
$\lambda\in I_a$ outside the eigenvalues of $H$.
\end{theorem}

\proof We first show that \eqref{eq:J} is satisfied for any open
interval $J$ whose closure is included in $I_a$, i.e.\
$\inf\{k\cdot u_N(k) \mid k\in X, |k|^2\in J\}>0$. Since $k\mapsto k\cdot u_N(k)$ is a
continuous function and $\bar{J}$ is a compact in $I_a$, it suffices to
check that $ak\cdot u_N(k)=\sum ak_j\sin(ak_j)>0$ for all $k$ such that
$|k|^2\in I_a$. The last condition may be written $0<|ak|<\pi$ and this
implies $|ak_j|<\pi$ for all $j$ and $|ak_j|>0$ for at least one
$j$. Clearly then we get $ak\cdot u_N(k)>0$.

For the rest of the proof it suffices to check that $V$ is of class
$C^2(A_N,\ch^1,\ch^{-1})$. Indeed, then we may use Theorems \ref{th:756}, Corollary
\ref{co:756s}, and Theorem \ref{th:reg} (see also Corollary
\ref{co:regm}).


Thus we have to prove that the commutators $[A_N,V]$ and $[A_N,[A_N,V]]$,
which are a priori defined as sesquilinear forms on
$C_\rmc^\infty(X)$, extend to continuous forms on $\ch^1$. Although
the computations are very simple, we give the details for the
convenience of the reader.

We have $[A_N,V]=\sum[A_j,V]$ and $[A_N,[A_N,V]]=\sum [A_j,[A_k,V]]$ and we
recall the relations \eqref{eq:aux}.
Since $\sin(ap_j)=\frac{1}{2\i}(T_j-T_j^*)$ and
$\cos(ap_j)=\frac{1}{2}(T_j+T_j^*)$ we have
\begin{equation}\label{eq:A0}
2\rmi A_j=q_j2\rmi\sin(ap_j)+ a \cos(ap_j)
=q_j (T_j-T_j^*) +b(T_j+T_j^*)
\end{equation}
where $b=a/2$.  Thus by using the relations $[T_j,V]=\delta_j(V)T_j$
and $[T^*_j,V]=-T_j^*\delta_j(V)$ and since $T_jq_jT_j^*=q_j+a$, we
get
\begin{eqnarray}\label{eq:A1}
[2\rmi A_j,V] &=& q_j[T_j-T_j^*,V]  +b[T_j+T_j^*,V] \nonumber\\
&=&(b+q_j) \delta_j(V)T_j + T_j^* (-b+q_j)\delta_j(V)
\end{eqnarray}
Because $T_j=e^{\rmi a p_j}$ and $T_j^*=e^{-\rmi a p_j}$ are bounded,
by the assumption $\mathbf{(H)}$ the right hand side of this relation
is a bounded operator from $\ch^1$ to $\ch^{-1}$, hence $V$ is of
class $C^1(A_N,\ch^1,\ch^{-1})$. It remains to treat the second order commutators.\\
Since $[\rmi A_N,S^*]=[\rmi A_N,S]^*$, we have
\begin{align}\label{eq:A1'}
[\rmi A_j,[\rmi A_k,V]] 
&=  [\rmi A_j, (b+q_k)\delta_k(V)T_k]+[\rmi A_j, T_k^* (-b+q_k)\delta_k(V)]\nonumber \\
&=[\rmi A_j, (b+q_k)\delta_k(V)]T_k+(b+q_k)\delta_k(V)[\rmi A_j, T_k]\nonumber\\
&+[\rmi A_j, T_k ]^*(-b+q_k)\delta_k(V)+T_k^*[\rmi A_j,  (-b+q_k)\delta_k(V)].
\end{align}
Since we have $[\rmi A_j, T_k]=b\delta_{jk} (1-T_k^2)$, we get
\begin{multline}\label{eq:A2}
[\rmi A_j, (b+q_k)\delta_k(V)]T_k+(b+q_k)\delta_k(V)[\rmi A_j, T_k]
= [\rmi A_j, (b+q_k)\delta_k(V)]T_k\\
+b\delta_{jk}  (b+q_k)\delta_k(V) (1-T_k^2) .
 \end{multline}
 The last term here is again a bounded operator $\ch^1\to\ch^{-1}$ by
 assumption $\mathbf{(H)}$, hence it remains to prove that the first
 term of the right hand side has the same property. For this we use
 \eqref{eq:A1} with $(b+q_k)\delta_k(V)$ instead of $V$ and we get:
\begin{multline}\label{eq:A3}
[\rmi A_j, (b+q_k)\delta_k(V)]=
 (b+q_j) \delta_j((b+q_k)\delta_k(V))T_j \\ 
 +T_j^* (-b+q_j)\delta_j((b+q_k)\delta_k(V)) .
\end{multline}
Since $\delta_j(MN)=\delta_j(M)T_jNT_j^*+M\delta_j(N)$ we have
\begin{align*}
\delta_j \big((b+q_k)\delta_k(V) \big) &=
\delta_j(b+q_k) T_j \delta_k(V) T_j^* +(b+q_k)\delta_j\delta_k(V) \\
&=a\delta_{jk} T_j \delta_j(V) T_j^* +(b+q_k)\delta_j\delta_k(V) .
\end{align*}
Since $(b+q_j)T_j=T_j(q_j-b)$ we then get  
\[
(b+q_j) \delta_j \big((b+q_k)\delta_k(V) \big)=
a\delta_{jk} T_j  (q_j-b)  \delta_j(V) T_j^* 
+(b+q_j)  (b+q_k)\delta_j\delta_k(V)
\]
which is bounded as operator $\ch^1\to\ch^{-1}$ by $\mathbf{(H)}$,
hence the right hand side of \eqref{eq:A3} has the same property. 

Using the same argument, we can prove that 
\[[\rmi A_j, T_k ]^*(-b+q_k)\delta_k(V)+T_k^*[\rmi A_j,  (-b+q_k)\delta_k(V)]\]
 is bounded. \\
 By \eqref{eq:A1'}, $[\rmi A_j,[\rmi A_k,V]] $ is bounded and we deduce that $V$ is of class $C^2(A_N,\ch^1,\ch^{-1})$.
\qed

\begin{remark}
This Theorem is a stronger version of Nakamura's result. In fact, in Nakamura's paper, $V$ is a multiplication operator with the $\Delta$-compact property (compact from $\ch^2$ to $L^2$) which is a stronger assumption than compact from $\ch^1$ to $\ch^{-1}$. Moreover, we add in Theorem \ref{th:nak} a result concerning the regularity of the boundary values of the resolvent. 
\end{remark}

%
%
%
%
%
Now we assume that $V$ and $S$ are real functions and we give more
explicit conditions which ensure that the assumptions of the Theorems
\ref{th:LAP-intro} and \ref{th:scatt} are satisfied. Let $p=1$ if
$\nu=1$, any $p>1$ if $\nu=2$, and $p=\nu/2$ if $\nu\geq 3$ and denote
\[
\llbracket f\rrbracket_p^x=\left({\textstyle\int_{|y-x|<1}}|f(y)|^p
\d y\right)^{1/p} . 
\] 
\begin{proposition}
Consider $V$ and $S$ multiplication operators such that:
\begin{itemize}
\item $V\in L^p_{\mathrm{loc}}(X)$ satisfies
  $\lim_{x\to\infty} \llbracket V\rrbracket_p^x =0$ and for any
  $a\in X$ and any $r>1$,
  \[\int_1^\infty\varphi_a(r) \frac{\d r}{r} <\infty\]
  where
  \[\varphi_a(r)=\sup_{|x|>r}\biggl\{|x|\llbracket V(\cdot+a) -V(\cdot)\rrbracket_{p}^x\biggr\};\]

\item $S\in L^p_{\mathrm{loc}}(X)$ and
  $\int_1^\infty \sup_{|x|>r}\llbracket S \rrbracket_p^x \, \d r<\infty$.

\end{itemize}
Then all the conditions of Theorems \ref{th:LAP-intro} and
\ref{th:scatt} are satisfied.
\end{proposition}
\proof
If
$U:\ch^1\to\ch$ is a local operator then it is easy to see that there
is $C'\in\R$ such that
\[
\|U\|_{\ch^1\to\ch}\leq C'\sup\{\|Uf\| \mid 
f\in\ch^1 \text{ with } \|f\|_{\ch^1}\leq1 \text{ and }
\mathrm{diam}\,\supp f \leq1 \}.
\]
If $V$ is a function and $\nu\geq3$ then the Sobolev inequality gives
a number $C''$ such that
\[
|\braket{f}{Vf}|\leq \||V|^{1/2}f\|^2 \leq 
\|V\|_{L^{\frac{\nu}{2}}} \|f\|^2_{L^\frac{2\nu}{\nu-2}}
\leq C'' \|V\|_{L^{\frac{\nu}{2}}} \|f\|^2_{\ch^1} \,.
\] 
If $\nu=1,2$ then the argument is simpler but $\nu/2$ has to be
replaced by $1$ or any $p>1$ respectively. Thus, if we introduce the
notation
\begin{equation}\label{eq:pnorm}
\llbracket V \rrbracket_p =
\sup_x\left({\textstyle\int_{|y-x|<1}}|V(y)|^p\d y \right)^{1/p}
\end{equation} 
with $p=1$ if $\nu=1$, any $p>1$ if $\nu=2$, and $p=\nu/2$ if
$\nu\geq3$, we get the following estimate: there is a number
$C=C(\nu,p)$ such that 
\begin{equation}\label{eq:sob}
\|V\|_{\rb}\leq C \llbracket V \rrbracket_p \,.
\end{equation}
Clearly that $C_\rmc^\infty(X)$ is dense for the norm
$\llbracket\cdot\rrbracket_p$ in the space of functions $V$ with
finite $\llbracket\cdot\rrbracket_p$ norm and such that
$\int_{|y-x|<1}|V(y)|^p\d y\to0$ as $x\to\infty$. Thus for such
functions the operator $V(q):\ch^1\to\ch^{-1}$ is compact.

Suppose that, $\varphi_a(r)=\sup_{|x|>r}\{|x|\llbracket V(\cdot+a) -V(\cdot)\rrbracket_{p}^x\}$ verifies
  
  \[\int_1^\infty\varphi_a(r) \frac{\d r}{r} <\infty.\]

 We will prove that $V$ verifies \eqref{eq:LR-intro}.
  
Because $\xi(x)=0$ if $\|x\|\leq 1$, and according to \eqref{eq:sob}, we have
  
\begin{equation}
\|\xi(q/r)|q| (V(q+a e) -V(q))\|_{\rb} \leq C \sup_{|x|>r}\llbracket |\cdot|(V(\cdot+a) -V(\cdot))\rrbracket_{p}^x .
\end{equation}

By definition, we have
\begin{equation}\label{eq:maj crochet}
\llbracket |\cdot|(V(\cdot+a) -V(\cdot))\rrbracket_{p}^x=\biggl(\int_{|y-x|<1}|y|^p|V(y+a) -V(y)|^p\d y\biggr)^{1/p}.
\end{equation}

Because $p\geq1$, $|y|^p\leq(|y-x|+|x|)^p$.
Moreover, using the convexity of the function $x\mapsto x^p$ on $\R^+$,
\begin{eqnarray}
(|y-x|+|x|)^p &=& 2^p\biggl(\frac{|y-x|+|x|}{2}\biggr)^p\nonumber\\
&\leq&2^{p-1}(|y-x|^p+|x|^p).\nonumber
\end{eqnarray}
So from \eqref{eq:maj crochet}, we have
\begin{equation}
\llbracket |\cdot|(V(\cdot+a) -V(\cdot))\rrbracket_{p}^x \leq C_2(1+|x|^p)^{1/p}\llbracket (V(\cdot+a) -V(\cdot))\rrbracket_{p}^x
\end{equation}
where $C_i$ are constants independent of $x$

By hypothesis on $V$, we have the following
\begin{equation}
\|\xi(q/r)|q| (V(q+a e) -V(q))\|_{\rb}\leq C_3\sup_{|x|>r}\frac{(1+|x|^p)^{1/p}}{|x|}\varphi_a(r)\leq C_4\varphi_a(r)
\end{equation}
and then
\begin{equation}
 \int_1^\infty \|\xi(q/r)|q| (V(q+a e) -V(q))\|_{\rb} 
  \frac{\d r}{r}<C_4 \int_1^\infty \varphi_a(r)
  \frac{\d r}{r} <\infty.
\end{equation}
\qed

A class of potentials that we may consider is the \emph{Kato
  class} whose definition is as follows \cite[Sec.\ 1.2]{CFKS}.  A
measurable function $V:\R^\nu\to\R$ is of class $K_\nu$ if
\begin{itemize}
\item 
$\lim_{\alpha\downarrow0}\sup_x\int_{|y-x|<\alpha} |y-x|^{2-\nu}|V(y)|\d
y=0$ in case $\nu>2$,
\item
$\lim_{\alpha\downarrow0}\sup_x\int_{|y-x|<\alpha}
\ln|y-x|^{-1}|V(y)|\d y =0$ in case $\nu=2$,
\item
$\lim_{\alpha\downarrow0}\sup_x\int_{|y-x|<\alpha} |V(y)|\d y =0$ in case $\nu=1$.
\end{itemize} 

The $K_\nu$ norm of such a function is given by 
\(
\|V\|_{K_\nu}=\sup_x\int_{|y-x|<1} L_\nu(y-x)|V(y)| \d y
\)
with the obvious definition of $L_\nu$. Note that the operator $V(q)$
is form relatively bounded with respect to the Laplacian with relative
bound zero if $V\in K_\nu$ \cite[p.\ 8]{CFKS} hence $H=\Delta+V(q)$ is
a well-defined self-adjoint and bounded from below operator.

\begin{proposition}\label{pr:Kato}
Let $V$ be a real function on $\R^\nu$, with $\nu\not=2$, such that there is $\mu>1$ and $\bigl(\langle\cdot\rangle^\mu V(\cdot)\bigr)^p\in K_\nu$.
Then condition \eqref{eq:LR-intro} is satisfied.
\end{proposition}

\proof
According to \eqref{eq:sob}, there is $C>0$ such that
\begin{eqnarray}\label{eq:Kato 1}
\|\langle q\rangle^\mu V\|_\rb&\leq& C\llbracket \langle \cdot\rangle^\mu V\rrbracket_p \,\nonumber\\
&\leq& C \sup_x\left({\textstyle\int_{|y-x|<1}}|\langle y\rangle^\mu V(y)|^p\d y \right)^{1/p}\nonumber\\
&\leq& C \left(\sup_x{\textstyle\int_{|y-x|<1}}|\langle y\rangle^\mu V(y)|^p\d y \right)^{1/p}
\end{eqnarray}

Because, if $\nu\not=2$, $L_\nu(y-x)\geq 1$ if $|y-x|<1$, we have the following

\begin{equation}
{\textstyle\int_{|y-x|<1}}|\langle y\rangle^\mu V(y)|^p\d y\leq{\textstyle\int_{|y-x|<1}} L_\nu(y-x)|\langle y\rangle^\mu V(y)|^p \d y 
\end{equation}

So
\begin{equation}
\sup_x{\textstyle\int_{|y-x|<1}}|\langle y\rangle^\mu V(y)|^p\d y\leq\sup_x{\textstyle\int_{|y-x|<1}} L_\nu(y-x)|\langle y\rangle^\mu V(y)|^p \d y
\end{equation}

According to \eqref{eq:Kato 1}, we have
\begin{equation}
\|\langle q\rangle^\mu V\|_\rb\leq C\left(\|(\langle \cdot\rangle^\mu V(\cdot))^p\|_{K_\nu}\right)^{1/p}.
\end{equation}
So if there is $\mu>1$ such that $\bigl(\langle\cdot\rangle^\mu V(\cdot)\bigr)^p\in K_\nu$, then $\|\langle q\rangle^\mu V\|_\rb<\infty$ and $V$ satisfies \eqref{eq:LR-intro}.
\qed




\section{Concrete potentials}\label{s:potential}

In this section, we will give examples of concrete potential which satisfy the assumptions of Theorem \ref{th:LAP-intro} and the assumptions of Theorem \ref{th:756}. For these examples, we will discuss the application of the Mourre Theorem with the generator of dilation and/or Nakamura's result.

Note that since $H_0=\Delta:\ch^1\rightarrow\ch^{-1}$ is bounded, if $V:\ch^1\rightarrow\ch^{-1}$ is compact, then $H:\ch^1\rightarrow\ch^{-1}$ is bounded. Furthermore $H_0\in C^\infty(A_u,\ch^1,\ch^{-1})$ and we can deduce that:

\begin{proposition}
Let $k\in\N^*$.
We suppose that $V:\ch^1\rightarrow\ch^{-1}$ is compact.
The folowing properties are equivalent:
\begin{enumerate}

\item $H=\Delta+V \in C^k(A_u)$;

\smallskip

\item $H=\Delta+V \in C^k(A_u,\ch^1,\ch^{-1})$;

\smallskip

\item $V\in C^k(A_u,\ch^1,\ch^{-1})$.
\end{enumerate}
\end{proposition} 

When $V:\ch^2\rightarrow L^2$ is compact, we have the following
\begin{proposition}
Let $k\in\N^*$.
We suppose that $V$ is $\Delta$-compact.
The folowing properties are equivalent:
\begin{enumerate}
\item $H=\Delta+V \in C^k(A_u,\ch^2,\ch^{-2})$;

\smallskip

\item $V\in C^k(A_u,\ch^2,\ch^{-2})$.
\end{enumerate}
\end{proposition}
Remark that if $V$ is $\Delta$-compact and $k\not=1$, $H\in C^k(A_u,\ch^2,\ch^{-2})$ is not equivalent to $H \in C^k(A_u)$ (see \cite[Theorem 6.3.4]{ABG}).

\subsection{A non Laplacian-compact potential}

In this part, we work in one dimension.\\
Let $\chi\in C^1(\R,\R)$ such that $\chi(x)=0$ if $|x|>1$, $\chi(x)>0$ if $|x|<1$ and $\chi(-x)=\chi(x)$.
\begin{lemma}\label{l:Nak regularity}
Let $V$ such that 
\[\widehat{qV}(\xi)=\sum_{n=-\infty}^{+\infty} \lambda_n \chi(\xi-n),\]
 where $\lambda_{-n}=\lambda_n\geq 0$, $\lambda_0=0$, and $(\lambda_n)_{n\in\Z}$ is not bounded. Moreover, we suppose that there is $0<\epsilon<1/2$ such that 
 \begin{equation}\label{eq:prop lambda
 }\sum_{n=-\infty}^{+\infty} \lambda_n\langle n\rangle^{-1/2+\epsilon}<\infty.
 \end{equation}
  Then, for all $u\in\cu$,
\begin{enumerate}
\item $V$ is symmetric and $V:\ch^1\rightarrow\ch^{-1}$ is compact.

\item $V\in C^{1,1}(A_u,\ch^1,\ch^{-1})$.

\item $V\notin C^1(A_D,\ch^1,\ch^{-1})$.

\item $V$ is not $\Delta$-bounded.
\end{enumerate}
 \end{lemma}
 We will give few remarks about this lemma.
 \begin{enumerate}[(a)]
 \item Note that, since $\chi$ is compactly support, the sum which defines $V$ is locally finite and so $V$ is well defined.
 
 \smallskip
 
\item Lemma \ref{l:Nak regularity} applies with $A_N$ replacing $A_u$ but since $V$ is not $\Delta$-bounded, and so $V$ is not $\Delta$-compact, we can not apply \cite[Theorem 1]{Nak}. Furthermore, because of (3), we can not apply the Mourre Theorem with $A_D$ as conjugate operator.

\smallskip

 \item The requirements on $(\lambda_n)_{n\in\Z}$ are satisfied in the case
 \[\lambda_n=\begin{cases}
p\quad\text{ if }|n|=2^p\\
0\quad\text{ else}
\end{cases}.\]
\end{enumerate}

\proof[Lemma \ref{l:Nak regularity}] \[\]
\begin{enumerate}
\item Let \[T_n(x)=\begin{cases}
\int_0^x\chi(s-n)ds \quad \text{ if } n>0\\
-\int_x^0\chi(s-n)ds \quad \text{ if } n<0
\end{cases}.\]
Remark that $T_n'=\chi(\cdot-n)$.
For $n>0$ and $f,g\in\ch^1$, we have
\begin{eqnarray*}
\left|(\hat{f},T_n*\hat{g})\right|&=&\left|\int_{\R^2}\hat{f}(\xi)\hat{g}(\eta)\int_0^{\xi-\eta}\chi(s-n)dsd\xi d\eta\right|\\
&=&\left|\int_{\R^2}\hat{f}(\xi)\hat{g}(\eta)\int_{-n}^{\xi-\eta-n}\chi(s)dsd\xi d\eta\right|.
\end{eqnarray*}
Since $\chi(s)=0$ if $s\leq-1$, $\int_{-n}^{\xi-\eta-n}\chi(s)ds=0$ if $\xi-\eta-n\leq-1$. If $\xi-\eta-n>-1$, $\int_{-n}^{\xi-\eta-n}\chi(s)ds=\int_{-1}^{\xi-\eta-n}\chi(s)ds$.
So
\begin{eqnarray*}
\left|(\hat{f},T_n*\hat{g})\right|&=&\left|\int_{\R^2}\hat{f}(\xi)\hat{g}(\eta)\int_{-1}^{\xi-\eta-n}\chi(s)dsd\xi d\eta\right|\\
&\leq&\int_{\R^2}\langle\xi\rangle|\hat{f}(\xi)|\langle\eta\rangle|\hat{g}(\eta)| \langle\xi\rangle^{-1}\langle\eta\rangle^{-1}\int_{-1}^{\xi-\eta-n}\chi(s)dsd\xi d\eta.
\end{eqnarray*}
Since $\epsilon<1/2$, there is $C>0$ such that $\langle\xi\rangle^{-1/2+\epsilon}\langle\eta\rangle^{-1/2+\epsilon}\leq C\langle\xi-\eta\rangle^{-1/2+\epsilon}$.
\begin{eqnarray*}
\left|(\hat{f},T_n*\hat{g})\right|&\leq&C \int_{\R^2}\langle\xi\rangle|\hat{f}(\xi)|\langle\eta\rangle|\hat{g}(\eta)| \langle\xi\rangle^{-1/2-\epsilon}\langle\eta\rangle^{-1/2-\epsilon}\\
& &\langle\xi-\eta\rangle^{-1/2+\epsilon}\int_{-1}^{\xi-\eta-n}\chi(s)dsd\xi d\eta\\
&\leq&\langle n-1\rangle^{-1/2+\epsilon}\int_{-1}^{1}\chi(s)ds\int_{\R^2}\langle\xi\rangle|\hat{f}(\xi)|\langle\eta\rangle|\hat{g}(\eta)| \\
& &\langle\xi\rangle^{-1/2-\epsilon}\langle\eta\rangle^{-1/2-\epsilon}d\xi d\eta.
\end{eqnarray*}
So, since $K:(\xi,\eta)\rightarrow \langle\xi\rangle^{-1/2-\epsilon}\langle\eta\rangle^{-1/2-\epsilon}$ is in $L^2(\R^2)$, the operator $L^2\ni \psi\mapsto \int_\R K(\xi,\eta)\psi(\eta)d\eta$ is compact (see \cite[Theorem VI.23]{RS1}). So $T_n$ is compact from $\ch^1$ to $\ch^{-1}$ if $n>0$ and we can remark that we have similar inequalities for $n<0$.

So
\begin{eqnarray*}
\left|(\hat{f},\hat{V}*\hat{g})\right|&\leq&\sum_{n=-\infty}^{+\infty}|\lambda_n|\left|(\hat{f},T_n*\hat{g})\right|\\
&\leq&\int_{\R^2}\langle\xi\rangle|\hat{f}(\xi)|\langle\eta\rangle|\hat{g}(\eta)| K(\xi,\eta)d\xi d\eta\\
& &\int_{-1}^{1}\chi(s)ds\sum_{n=-\infty}^{+\infty}|\lambda_n|\max(\langle n-1\rangle^{-1/2+\epsilon},\langle n+1\rangle^{-1/2+\epsilon}).
\end{eqnarray*}

So, since $\sum_{n=-\infty}^{+\infty}|\lambda_n|\langle n\rangle^{-1/2+\epsilon}<\infty$, $V:\ch^1\rightarrow\ch^{-1}$ is compact. 

Moreover, since $\chi(-x)=\chi(x)$ and $\lambda_n=\lambda_{-n}$, we have
\begin{eqnarray*}
\widehat{qV}(-\xi)&=&\sum_{n=-\infty}^{+\infty} \lambda_n \chi(-\xi-n)\\
&=&\sum_{n=-\infty}^{+\infty} \lambda_n \chi(\xi+n)\\
&=&\sum_{n=-\infty}^{+\infty} \lambda_{-n} \chi(\xi-n)\\
&=&\widehat{qV}(\xi)
\end{eqnarray*}
So, since $\widehat{qV}(\xi)\in\R$, $xV(x)\in\R, \forall x\in\R$, and we conclude that $V$ is a symmetric multiplication operator.

\smallskip

\item By a simple computation, in all dimension, we have
\begin{eqnarray}\label{eq:commutateur V,A}
(g,[V,A_u]f)&=&(Vg,A_u f)-(A_ug,Vf)\nonumber\\
&=&(Vg,qu(p)f)-(qu(p)g,Vf)\nonumber\\
& &-\frac{i}{2}\left((Vg,u'(p)f)+(u'(p)g,Vf)\right)\nonumber\\
&=&(qVg,u(p)f)-(u(p)g,qVf)\nonumber\\
& &\qquad-\frac{i}{2}\left((Vg,u'(p)f)+(u'(p)g,Vf)\right).
\end{eqnarray}

If $u'$ is bounded, since $V:\ch^1\rightarrow\ch^{-1}$ is a compact operator, there is $C>0$ such that
\[\left|(Vg,u'(p)f)+(u'(p)g,Vf)\right|\leq C \|f\|_{\ch^1}\|g\|_{\ch^{1}}.\]

Moreover, we can remark that $\langle \cdot\rangle^{-1}\widehat{qV}\in L^1(\R)$. So, there is $C>0$ such that
\begin{eqnarray}\label{eq:norme H1}
|(u(p) f,qV g)|&=&|(u(q) \hat{f},\widehat{qV}*\hat{g})|\nonumber\\
&=&\left|\int_{\R^2} u(\xi)\hat{f}(\xi)\widehat{qV}(\xi-\eta)\hat{g}(\eta)d\xi d\eta\right|\nonumber\\
&\leq&C\int_{\R^2} |u(\xi)|\langle \xi\rangle|\hat{f}(\xi)|\langle \xi-\eta\rangle^{-1}|\widehat{qV}(\xi-\eta)|\nonumber\\
& &\qquad \langle \eta\rangle|\hat{g}(\eta)|d\xi d\eta
\end{eqnarray}
Since $f,g\in\ch^1$ and $u$ is bounded, $u(q)\langle q\rangle\hat{f}$ and $\langle q\rangle\hat{g}$ are in $L^2$. So by Young inequality, we conclude that
\begin{eqnarray*}
|(u(p) f,qV g)|&\leq& C\|u(q)\langle q\rangle\hat{f}\|_2\|(\langle q\rangle^{-1}\widehat{qV})*(\langle q\rangle\hat{g})\|_2\\
&\leq& C \|\langle q\rangle^{-1}\widehat{qV})\|_1\|\langle q\rangle\hat{f}\|_2\|\langle q\rangle\hat{g}\|_2\\
&\leq& C \|\langle q\rangle^{-1}\widehat{qV})\|_1\|f\|_{\ch^1}\|g\|_{\ch^1}.
\end{eqnarray*}
So $V\in C^1(A_u,\ch^1,\ch^{-1})$.
Similarly, we have
\begin{eqnarray*}
\left(g,[[V,A_u],A_u]f\right)&=&\left([V,A_u]g,A_uf\right)-\left(A_ug,[V,A_u]f\right)\\
&=&\left([V,A_u]g,qu(p)f\right)-\left(qu(p)g,[V,A_u]f\right)\\
& &-\frac{i}{2}\biggl(\left(u'(p)g,[V,A_u]f\right)+\left([V,A_u]g,u'(p)f\right)\biggr).
\end{eqnarray*}
Since $V\in C^1(A_u,\ch^1,\ch^{-1})$, $\left(u'(p)g,[V,A_u]f\right)+\left([V,A_u]g,u'(p)f\right)$ is bounded. Using \eqref{eq:commutateur V,A}, we have 
\begin{eqnarray*}
\left(qu(p)g,[V,A_u]f\right)&=&\left(qVqu(p)g,u(p)f\right)-\left(u(p)qu(p)g,qVf\right)\\
& &-\frac{i}{2}\biggl(\left(Vqu(p)g,u'(p)f\right)+\left(u'(p)qu(p)g,Vf\right)\biggr)\\
&=&\left(q^2Vu(p)g,u(p)f\right)-\left(u(p)qu(p)g,qVf\right)\\
& &-\frac{i}{2}\biggl(\left(u(p)g,qVu'(p)f\right)+\left(u'(p)qu(p)g,Vf\right)\biggr).
\end{eqnarray*}
By a simple computation, we have
\begin{eqnarray*}
\left(u(p)qu(p)g,qVf\right)&=&\left(qu(p)u(p)g,qVf\right)+\left([u(p),q]u(p)g,qVf\right)\\
&=&\left(u(p)u(p)g,q^2Vf\right)+\frac{i}{2}\left(u(p)u'(p)g,qVf\right)
\end{eqnarray*}
and
\begin{eqnarray*}
\left(u'(p)qu(p)g,Vf\right)&=&\left(qu'(p)u(p)g,Vf\right)+\left([u'(p),q]u(p)g,Vf\right)\\
&=&\left(u(p)u'(p)g,qVf\right)+\frac{i}{2}\left(u''(p)u(p)g,Vf\right).
\end{eqnarray*}
As previously, we can remark that $\langle q\rangle^{-1} \widehat{q^2V}\in L^1$.
So, since $u$ and all of whose derivatives are bounded, we deduce that
 $[[V,A_u],A_u]$ is a bounded operator on $\ch^1\rightarrow\ch^{-1}$ and $V\in C^2(A_u,\ch^1,\ch^{-1})$. Remark that all previous inequalities are true for any bounded $u$ such that $u\in C^\infty$ with all derivatives bounded. In particular, by taking $u(x)=sin(ax)$, we deduce that \[V\in C^2(A_N,\ch^1,\ch^{-1})\subset  C^{1,1}(A_N,\ch^1,\ch^{-1}).\]

\smallskip
 
\item Now we will prove that $V$ is not in $C^1(A_D,\ch^1,\ch^{-1})$.\\
For $N\in\N^*$, let 
\[\widehat{f_N}=\mathbb{1}_{[N,N+1]}\langle N+1\rangle^{-1}\quad
\text{ and} \quad \hat{g}=\mathbb{1}_{[0,1]}.\] 
Note that $\|\langle\cdot\rangle\widehat{f_N}\|^2_{L^2}\|\langle\cdot\rangle\widehat{g}\|^2_{L^2}\leq 2$ which implies that $\|f_N\|_{\ch^1}\|g\|_{\ch^1}\leq 4\sqrt{2}$.\\
We have
\begin{eqnarray*}
\left(\widehat{f_N}, \widehat{\nabla(qV)}*\hat{g}\right)&=&\sum_{n=-\infty}^{+\infty} \lambda_n \int_{\R^2}(\xi-\eta)\widehat{f_N}(\xi)\chi(\xi-\eta-n)\hat{g}(\eta)d\xi d\eta\\
&\geq& \frac{\lambda_N}{\langle N+1\rangle}\int_N^{N+1}\left(\int_0^1(\xi-\eta)\chi(\xi-\eta-N+1)d\eta\right) d\xi\\
&\geq& \frac{\lambda_N}{\langle N+1\rangle}(N-1)\int_N^{N+1}\left(\int_{\xi-N}^{\xi-N+1}\chi(\sigma)d\sigma\right) d\xi\\
&\geq&\frac{N-1}{\langle N+1\rangle}\lambda_N\int_{0}^{2}\chi(\sigma)\left(\int_{N+\sigma-1}^{N+\sigma} d\xi\right)d\sigma \\
&\geq&\frac{N-1}{\langle N+1\rangle}\lambda_N\int_{0}^{2}\chi(\sigma)d\sigma.
\end{eqnarray*}
So, since $\int_{0}^{2}\chi(\sigma)d\sigma>0$, $\biggl((f_N,[V,iA_D]g)\biggr)_{N\in\N}$ is not bounded with \\
$\|f_N\|_{\ch^1}\|g\|_{\ch^1}\leq 4$.
 So $V$ does not belong to the class $C^1(A_D,\ch^1,\ch^{-1})$.
 
 \smallskip
 
\item Now, we will prove that $V$ is not $\Delta$-compact.\\
Let $N\in\N$, $N\geq2$ and let 
\[\widehat{f_N}=\mathbb{1}_{[N+1,N+2]}\quad \text{and} \quad\hat{g}=\mathbb{1}_{[0,1]}.\]
Remark that $\|f_N\|_{L^2}\|g\|_{\ch^2}$ is a bounded sequence.
 
 \begin{eqnarray}\label{eq:Delta compact}
 \left|(f_N,Vg)\right|&=&\left|(\widehat{f_N},\hat{V}*\hat{g})\right|\nonumber\\
 &=&\int_{\xi\in[N+1,N+2]}\int_{\eta\in[0,1]}\sum_{n=-\infty}^{+\infty}\lambda_n\int_0^{\xi-\eta}\chi(s-n)ds\nonumber\\
 &=&-\int_{\xi\in[N+1,N+2]}\int_{\eta\in[0,1]}\sum_{n=-\infty}^{-1}\lambda_n\int_{\xi-\eta}^0\chi(s-n)ds\nonumber\\
 & &+\int_{\xi\in[N+1,N+2]}\int_{\eta\in[0,1]}\sum_{n=1}^{N-1}\lambda_n\int_0^{\xi-\eta}\chi(s-n)ds\nonumber\\
 & &+\sum_{n=N}^{N+2}\lambda_n\int_{\xi\in[N+1,N+2]}\int_{\eta\in[0,1]}\int_0^{\xi-\eta}\chi(s-n)ds\nonumber\\
 & &+\int_{\xi\in[N+1,N+2]}\int_{\eta\in[0,1]}\sum_{n=N+3}^{+\infty}\lambda_n\int_0^{\xi-\eta}\chi(s-n)ds.
 \end{eqnarray}
 Remark that, for $\xi\in[N+1,N+2]$ and $\eta\in[0,1]$, $N+2\geq\xi-\eta\geq N$. So, if $\xi\in[N+1,N+2]$ and $\eta\in[0,1]$, we have:
 \begin{itemize}
 
\item If $n\leq -1$, $\chi(s-n)=0$ for all $s\in[0,\xi-\eta]$. So $\int_{\xi-\eta}^0\chi(s-n)ds=0$.
 
 \smallskip
 
\item If $1\leq n\leq N-1$, $\xi-\eta\geq n+1$.So, since $\chi(s)=0$ if $|s|\geq 1$, \[\int_0^{\xi-\eta}\chi(s-n)ds=\int_{n-1}^{n+1}\chi(s-n)ds=\int_{-1}^{1}\chi(s)ds>0.\]

\smallskip

\item If $ n\geq N+3$, $\xi-\eta\leq n-1$. So, since $\chi(s-n)=0$ for all $s\in[0,\xi-\eta]$, $\int_0^{\xi-\eta}\chi(s-n)ds=0$.
 \end{itemize}
 
 So, since $\lambda_n\geq0$ for all $n$ and $\chi(x)\geq0$ for all $x$, from \eqref{eq:Delta compact}, we have:
 \begin{eqnarray*}
  \left|(f_N,Vg)\right|&=&\int_{\xi\in[N+1,N+2]}\int_{\eta\in[0,1]}\sum_{n=1}^{N-1}\lambda_n\int_{-1}^{1}\chi(s)ds\\
 & &+\sum_{n=N}^{N+2}\lambda_n\int_{\xi\in[N+1,N+2]}\int_{\eta\in[0,1]}\int_0^{\xi-\eta}\chi(s-n)ds\\
 &\geq&\lambda_{N-1}\int_{-1}^{1}\chi(s)ds.
 \end{eqnarray*}
 
 So, since $(\lambda_n)_{n\in\N}$ is not bounded, we can extract a subsequence $(\lambda_{\phi(n)})_{n\in\N}$ such that 
 \[\lim\limits_{n\rightarrow+\infty}  \lambda_{\phi(n)}=+\infty\quad\text{ and we have }\quad \lim\limits_{N\rightarrow+\infty}\left|(f_{\phi(N)+1},Vg)\right|=+\infty.\] 
 So $V$ is not $\Delta$-bounded. \qed
 \end{enumerate}

\subsection{A class of oscillating potential}

Let $\alpha>0$, $\beta\in\R$, $k\in\R^*$ and $\kappa\in C^\infty_c(\R,\R)$ such that $\kappa=1$ on $[-1,1]$ and $0\leq \kappa\leq 1$.
Let \begin{equation}\label{eq:oscillant}
W_{\alpha \beta}(x)=(1-\kappa(|x|))\frac{\sin(k|x|^\alpha)}{| x|^\beta}.
\end{equation}
This potential can be seen as a $\Delta$-compact potential (if $\beta>0$) or as a potential on $\ch^1$ to $\ch^{-1}$ for which we keep the same notation.

We will see that under certain condition on $(\alpha,\beta)$, we can apply Theorems \ref{th:LAP-intro} and \ref{th:756} with $W_{\alpha \beta}$ as potential. We will also compare our results (lemma \ref{l:oscillant}) with results given in \cite{JMb}. 
 
 Recall that $\cu$ is the space of vector fields $u$ bounded with all derivatives bounded such that $x\cdot u(x)>0$ for all $x\not=0$. We have the following:
\begin{lemma}\label{l:oscillant}
Let $W_{\alpha \beta}$ be as in \eqref{eq:oscillant} and let $H=\Delta+W_{\alpha \beta}$. For all $u\in\cu$, we have:
\begin{enumerate}
\item \label{itm: beta neg} if $\alpha+\beta>2$, then $W_{\alpha \beta}:\ch^1\rightarrow\ch^{-1}$ is compact and $W_{\alpha \beta}\in C^{1,1}(A_u,\ch^1,\ch^{-1})$.

\smallskip

\item if $2\alpha+\beta>3$ and $\beta>0$, $W_{\alpha \beta}\in C^{1,1}(A_u,\ch^2,\ch^{-2})$.

\end{enumerate}
In particular, in this both cases, Theorems \ref{th:LAP-intro} and \ref{th:756} apply.
\end{lemma}

Note that in \eqref{itm: beta neg}, we do not require to have $\beta>0$. In particular, if $\beta<0$, $W_{\alpha \beta}$ is an unbounded function.

In \cite{JMb}, if we suppose $\beta>0$, we can see that the Limiting Absorption Principle can be proved with the generator of dilation $A_D$ as conjugate operator for $H=\Delta+W_{\alpha \beta}$ if $|\alpha-1|+\beta>1$. If $|\alpha-1|+\beta<1$, they showed that $H\notin C^1(A_D)$. This implies that we cannot apply the Mourre Theorem with $A_D$ as conjugate operator on this area. Moreover, they also proved a limiting absorption principle if $\alpha>1$ and $\beta>1/2$, in a certain energy window. If $|\alpha-1|+\beta<1$ and $2\alpha+\beta>3$, Theorem \ref{th:756} improves this result in two waves: first, there is no restriction of energy; second, we have some result on the boundary value of the resolvent. Furthermore, the region where $|\alpha-1|+\beta<1$, $2\alpha+\beta>3$ and $\beta\leq\frac{1}{2}$ is not covered by \cite{JMb} but Theorem \ref{th:756} applies.

\proof[Lemma \ref{l:oscillant}]
Let $f,g\in\cs$ and let $0<\mu<1$. Let $u\in\ru$. We will always suppose that $\mu$ is small enough. We have
\begin{eqnarray}
& &(f,\jap{q}^\mu[W_{\alpha \beta},iA_u]g)\nonumber\\
&=&(\jap{q}^\mu W_{\alpha \beta}f,iA_ug)-(A_u\jap{q}^\mu f,iW_{\alpha \beta}g)\nonumber\\
&=& (\jap{q}^\mu W_{\alpha \beta}f,iq\cdot u(p)g)-(q\cdot u(p)\jap{q}^\mu f,iW_{\alpha \beta}g)\nonumber\\
& &+(\jap{q}^\mu W_{\alpha \beta}f,\frac{1}{2}u'(p)g)+(\frac{i}{2}u'(p)\jap{q}^\mu f,iW_{\alpha \beta}g)\nonumber\\
&=&(\jap{q}^\mu W_{\alpha \beta}f,iq\cdot u(p)g)-(q\cdot u(p) f,i\jap{q}^\mu W_{\alpha \beta}g)\label{eq: com oscillant 1}\\
& &+(\jap{q}^\mu W_{\alpha \beta}f,\frac{1}{2}u'(p)g)+(\frac{i}{2}u'(p) f,i\jap{q}^\mu W_{\alpha \beta}g)\label{eq: com oscillant 2}\\
& &-(q\cdot [u(p),\jap{q}^\mu] f,iW_{\alpha \beta}g)+(\frac{i}{2}[u'(p),\jap{q}^\mu] f,iW_{\alpha \beta}g).\label{eq: com oscillant 3}
\end{eqnarray}
Remark that, since $u$ and all its derivatives are bounded, for $\mu<1$, $[u(p),\jap{q}^\mu]$ and $[u'(p),\jap{q}^\mu]$ are bounded and $\|u(p)f\|_{\ch^s}$ and $\|u'(p)f\|_{\ch^s}$ are controlled by $\|f\|_{\ch^s}$ for $s=1,2$. We will use this argument to treat terms in \eqref{eq: com oscillant 2} and \eqref{eq: com oscillant 3} and we note that they are bounded in the $\ch^1$ norm when terms in \eqref{eq: com oscillant 1} are bounded. For this reason, we focus on this to terms which are quite similar. To control them, we will show that $qW_{\alpha \beta}(q)$ can be write with a different form. 

Let $\tilde{\kappa}\in C^\infty_c(\R,\R)$ such that $\tilde{\kappa}(|x|)=0$ if $|x|\geq1$, $\tilde{\kappa}=1$ on $[-1/2,1/2]$ and $0\leq \tilde{\kappa}\leq 1$. So, we can observe that $(1-\tilde{\kappa}(|x|))(1-\kappa(|x|))=(1-\kappa(|x|))$ for all $x\in\R^\nu$.

For $\gamma\in\R$, let 

\[\tilde{W}_{\alpha \gamma}(x)=(1-\tilde{\kappa}(|x|))\frac{\cos(k|x|^\alpha)}{| x|^\gamma}.\]

By a simple computation, we have
\begin{equation}\label{eq:1grad}
(1-\kappa(|x|))|x|\nabla\tilde{W}_{\alpha \gamma}(x)=-(1-\kappa(|x|))\gamma\frac{x}{|x|}\tilde{W}_{\alpha \gamma}(x)
-k\alpha xW_{\alpha\beta}(x)
\end{equation}
with $\gamma=\alpha+\beta-1$.

In a first time, remark that, since, in both cases, $\gamma>0$,
 \[\langle q\rangle^\mu\frac{\gamma} {k\alpha}\frac{q}{| q|}\tilde{W}_{\alpha \gamma}(q)\]
 is bounded for all $0<\mu\leq\gamma$. Thus, by using \eqref{eq:1grad} in \eqref{eq: com oscillant 1}, it suffices to proof that 
 \[\jap{q}^\mu(1-\kappa(|q|))|q|\nabla\tilde{W}_{\alpha \gamma}(q):\ch^1\rightarrow\ch^{-1}\]
  is bounded to show that $W_{\alpha\beta}\in C^{1,1}(A_u,\ch^1,\ch^{-1})$.
 
To do this, remark that, for all function $F$, $\nabla F(q)=i[p,F(q)]$. So, we have for $\phi,\psi\in\cs$,
 \begin{eqnarray*}
& & (u(p)\phi,(1-\kappa(|q|))\jap{q}^\mu |q|\nabla\tilde{W}_{\alpha \gamma}(q)\psi)\\
&=&(u(p)\phi,i[p,(1-\kappa(|q|))\jap{q}^\mu |q|\tilde{W}_{\alpha \gamma}(q)]\psi)
 +(u(p)\phi,q\kappa'(|q|)\jap{q}^\mu \tilde{W}_{\alpha \gamma}(q)\psi)\\
& &-\mu(u(p)\phi,q(1-\kappa(|q|))\jap{q}^{\mu-1} |q|\tilde{W}_{\alpha \gamma}(q)\psi)\\
& &-(u(p)\phi,(1-\kappa(|q|))\jap{q}^\mu \frac{q}{|q|}\tilde{W}_{\alpha \gamma}(q)\psi).
 \end{eqnarray*}
 So, we have
  \begin{eqnarray}\label{eq:grad}
 & &(u(p)\phi,(1-\kappa(|q|))\jap{q}^\mu |q|\nabla\tilde{W}_{\alpha \gamma}(q)\psi)\nonumber\\
 &=&(pu(p)\phi,i(1-\kappa(|q|))\jap{q}^\mu |q|\tilde{W}_{\alpha \gamma}(q)\psi)\nonumber\\
& & -(u(p)\phi,i(1-\kappa(|q|))\jap{q}^\mu |q|\tilde{W}_{\alpha \gamma}(q)p\psi)+(u(p)\phi,q\kappa'(|q|)\jap{q}^\mu \tilde{W}_{\alpha \gamma}(q)\psi)\nonumber\\
 & &-\mu(u(p)\phi,q(1-\kappa(|q|))\jap{q}^{\mu-1} |q|\tilde{W}_{\alpha \gamma}(q)\psi)\nonumber\\
 & &-(u(p)\phi,(1-\kappa(|q|))\jap{q}^\mu \frac{q}{|q|}\tilde{W}_{\alpha \gamma}(q)\psi).
 \end{eqnarray}
 So, since $u$ is bounded, by density, if \\$(1-\kappa(|q|))\jap{q}^\mu |q|\tilde{W}_{\alpha \gamma}(q)$ is bounded, then $(1-\kappa(|q|))\jap{q}^\mu |q|\nabla\tilde{W}_{\alpha \gamma}(q):\ch^1\rightarrow\ch^{-1}$ is bounded.
 
 \begin{enumerate}
 \item Suppose that $\alpha+\beta>2$. Since $\gamma>1$, by \eqref{eq:1grad}, $\jap{q}^\mu qW_{\alpha\beta}:\ch^1\rightarrow\ch^{-1}$ is bounded for $\mu>0$ small enough. This implies that $W_{\alpha\beta}$ belongs to the class $C^{1,1}(A_u,\ch^1,\ch^{-1})$.
 
Moreover, by \eqref{eq:1grad}, we have
\[
W_{\alpha\beta}(x)=-(1-\kappa(|x|))\frac{\gamma} {k\alpha}\frac{1}{|x|}\tilde{W}_{\alpha \gamma}(x)
-(1-\kappa(|x|))\frac{x} {k\alpha|x|}\nabla\tilde{W}_{\alpha \gamma}(x).
\]
So, since $\gamma>0$, $(1-\kappa(|q|))\frac{\gamma} {k\alpha}\frac{1}{|q|}\tilde{W}_{\alpha \gamma}(q):\ch^1\rightarrow\ch^{-1}$ is compact. As in \eqref{eq:grad}, we can prove that $(1-\kappa(|q|))\frac{q} {k\alpha|q|}\nabla\tilde{W}_{\alpha \gamma}(q):\ch^1\rightarrow\ch^{-1}$ is compact. Thus, by sum, $W_{\alpha \beta}:\ch^1\rightarrow\ch^{-1}$ is compact.

\smallskip

\item Suppose that $\beta>0$ and $2\alpha+\beta>3$. In this case, remark that $W_{\alpha \beta}(q)$ is $\Delta$-compact. Let $u\in \cu$.  Let $\undertilde{\kappa}\in C^\infty_c(\R,\R)$ such that $\undertilde{\kappa}(x)=0$ if $|x|\geq1/2$, $\undertilde{\kappa}=1$ on $[-1/4,1/4]$ and $0\leq \undertilde{\kappa}\leq 1$. For $\delta\in\R$, let 

\[\undertilde{W}_{\alpha \delta}(x)=(1-\undertilde{\kappa}(|x|))\frac{\sin(k|x|^\alpha)}{| x|^\delta}.\]
By a simple computation, we can write:
\[
\tilde{W}_{\alpha\gamma}(x)=(1-\tilde{\kappa}(|x|))\frac{\delta} {k\alpha}\frac{1}{|x|}\undertilde{W}_{\alpha \delta}
+(1-\tilde{\kappa}(|x|))\frac{x} {k\alpha|x|}\nabla \undertilde{W}_{\alpha \delta}(x)
\]
with $\delta=\gamma+\alpha-1=\beta+2\alpha-2>1$.\\
Since we want to prove that $[W_{\alpha \beta},i A_u]:\ch^2\rightarrow\ch^{-2}$ is bounded, we can make twice the argument of \eqref{eq:grad}. This implies that $W_{\alpha \beta}$ is in $C^{1,1}(A_u,\ch^2,\ch^{-2})$.\qed
 \end{enumerate}
 
 If we want more regularity on the potentials, we have the following
 \begin{lemma}\label{l:reg sup}
 Let $V:\ch^1\rightarrow\ch^{-1}$ be a compact symmetric multiplication operator.
 If $|q|^n V:\ch^1\rightarrow\ch^{-1}$ is bounded for some $n\in\N^*$, then, for any $u\in\cu$, $V\in C^n(A_u,\ch^1,\ch^{-1})$.
 In particular, if $\alpha+\beta-1\geq n\in\N^*$, $W_{\alpha \beta}\in C^n(A_u,\ch^1,\ch^{-1})$. In this case, if $n>1$, we deduce that $\lambda\mapsto R(\lambda\pm i0)$ is locally of class $\Lambda^s$ on $\R^{+*}$ outside the eigenvalues of $H$, where $s$ is the integer part of $\alpha+\beta-2$.
 \end{lemma}
 \proof[Lemma \ref{l:reg sup}]
 Let $u\in\cu$ and  $n\in\N^*$. Let $ad^n_{A_u}(V)$ the iterated commutator of order $n$, with $ad^1_{A_u}(V)=[V,A_u]$.
By induction, we can prove that there is $\left(B_k(p)\right)_{k\in\{0,\cdots, n\}}$ and $\left(B'_k(p)\right)_{k\in\{0,\cdots, n\}}$ two sequences of bounded operators such that
\begin{equation}\label{eq: commutateur multiple}
 ad^n_{A_u}(V)=\sum_{k=0}^n B_k(p)q^kVB'_k(p).
 \end{equation}
 Moreover, we can see that $B_k(p)$ and $B'_k(p)$ depends only of $u$ and its derivatives of order less than $n$.
 So, if $|q|^n V:\ch^1\rightarrow\ch^{-1}$ is bounded, then $V\in C^n(A_u,\ch^1,\ch^{-1})$.
 
 Moreover, by \eqref{eq:1grad}, we can see that if $\gamma=\alpha+\beta-1\geq s+1, s\in\N^*$, then $|q|^{s+1} V:\ch^1\rightarrow\ch^{-1}$ is bounded. So $W_{\alpha \beta}\in C^{s+1}(A_u,\ch^1,\ch^{-1})\subset\Lambda^{s+1}(A_u,\ch^1,\ch^{-1})$ and by Theorem \ref{th:reg}, we can deduce that $\lambda\mapsto R(\lambda\pm i0)$ is locally of class $\Lambda^s$ on $\R^{+*}$ outside the eigenvalues of $H$. 
 \qed

\subsection{An unbounded potential with high oscillations}

Now, we will show an example of potential $V$ of class $C^\infty(A_u,\ch^1,\ch^{-1})$ for any $u\in\cu$ such that $V$ is neither in $C^1(A_D,\ch^1,\ch^{-1})$ nor $\Delta$-bounded. In particular we cannot have the Mourre estimate with $A_D$ as conjugate operator but we can prove a limiting absorption principle with $A_u$ as conjugate operator and have a good regularity for the boundary value of the resolvent.

\begin{lemma}\label{l:ex potentiel Cinfty}
Let $\kappa\in C^\infty_c(\R,\R)$ such that $\kappa=1$ on $[-1,1]$ and $0\leq \kappa\leq 1$.
Let
\begin{equation}\label{eq:ex expo}
V(x)=(1-\kappa(|x|))\exp(3|x|/4)\sin(\exp(|x|)).
\end{equation}
Then:
\begin{enumerate}
 \item $V:\ch^1\rightarrow\ch^{-1}$ is compact;
 
 \smallskip
 
\item For any $u\in\cu$, $V\in C^\infty(A_u,\ch^1,\ch^{-1})$;

\smallskip

\item $V$ is not $\Delta$-bounded;

\smallskip

\item $V$ is not of class $C^1(A_D,\ch^1,\ch^{-1})$.
\end{enumerate}
\end{lemma}
In particular, we can use neither the Mourre Theorem with the generator of dilation as conjugate operator nor Nakamura's Theorem. By Theorem \ref{th:reg}, we have the following
\begin{corollary}
Let $V$ as in \eqref{eq:ex expo} and $H=\Delta+V:\ch^1\rightarrow\ch^{-1}$.
Then Theorem \ref{th:756} applies and,
for all $s>0$, the functions
\begin{equation}
\lambda\mapsto R(\lambda\pm\i0)\in B(\ch^{-1}_s,\ch^1_{-s})
\end{equation}
are locally of class $\Lambda^{s-1/2}$ on $(0,+\infty)$ outside the eigenvalues
of $H$. 
\end{corollary}
In particular, if we see $R(\lambda\pm\i0)$ as an operator from $C^\infty_c$ to $\cd'$ the space of distributions, the functions
\[\lambda\mapsto R(\lambda\pm\i0)\in B(C^\infty_c,\cd')\]
are of class $C^\infty$ on $(0,+\infty)$ outside the eigenvalues.

\proof[Lemma \ref{l:ex potentiel Cinfty}]
Let $\tilde{\kappa}\in C^\infty_c(\R,\R)$ such that $\tilde{\kappa}(|x|)=0$ if $|x|\geq1$, $\tilde{\kappa}=1$ on $[-1/2,1/2]$ and $0\leq \tilde{\kappa}\leq 1$. So, we can observe that $(1-\tilde{\kappa}(|x|))(1-\kappa(|x|))=(1-\kappa(|x|))$ for all $x\in\R^\nu$.

If we denote 
\[\tilde{V}(x)=(1-\tilde{\kappa}(|x|))\cos(\exp(|x|)),\]
 we have:
\[(1-\kappa(|x|))\nabla\tilde{V}(x)=-(1-\kappa(|x|))\frac{x}{|x|}\exp(|x|)\sin(\exp(|x|)).\]
So,
\[xV(x)=-|x|(1-\kappa(|x|))\exp(-|x|/4)\nabla\tilde{V}(x).\]
\begin{enumerate}
\item By a simple calculus, we have 
\[V(x)=-\frac{x}{|x|}(1-\kappa(|x|))\exp(-|x|/4)\nabla\tilde{V}(x).\]
So, since $\tilde{V}$ is bounded, by writing $\nabla \tilde{V}(q)=i[p,\tilde{V}(q)]$, as in \eqref{eq:grad}, we can prove that $V:\ch^1\rightarrow\ch^{-1}$ is compact.

\smallskip

\item Similarly, since $\tilde{V}$ is bounded, by writing $\nabla \tilde{V}(q)=i[p,\tilde{V}(q)]$, \\$q|q|^nV(q):\ch^1\rightarrow\ch^{-1}$ is bounded for all $n\in\N$. So, by Lemma \ref{l:reg sup}, for any $u\in\cu$, $V\in C^n(A_u,\ch^1,\ch^{-1})$ for all $n\in\N^*$. 
So $V\in C^\infty(A_u,\ch^1,\ch^{-1})$.

\smallskip

\item Let $\chi\in C^0_c(\R,\R)$ such that $\supp(\chi)\subset [\frac{\pi}{4},\frac{3\pi}{4}]$, $\chi\geq0$ and $\chi(\pi/2)=1$.\\
Let $N\in\N^*$ and
\[f(x)=\jap{x}^{-(\nu+1)/2} \]
 and 
\[g_N(x)=\ln\left(\frac{3\pi}{4}+2N\pi\right)^{(1-\nu)/2}\exp(|x|/2)\chi\biggl(\exp(|x|)-2N\pi\biggr).\]
We denote $C>0$ constants independant of $N$.
Remark that $f\in\ch^2$ and 
\begin{eqnarray*}
\int_{\R^\nu} g_N^2(x)dx&=&\int_{\R^\nu} \ln\left(\frac{3\pi}{4}+2N\pi\right)^{(1-\nu)}\exp(|x|)\chi^2(\exp(|x|)-2N\pi)dx\\
&=& C\ln\left(\frac{3\pi}{4}+2N\pi\right)^{(1-\nu)}\int_{\R} \exp(r)\chi^2(\exp(r)-2N\pi)r^{\nu-1}dr\\
&\leq& C \int_{\R} \exp(r)\chi^2(\exp(r)-2N\pi)dr.
\end{eqnarray*}
So, if $\sigma=e^r-2N\pi$, we have
\begin{equation}\label{eq: g_N 2}
\int_{\R^\nu} g_N^2(x)dx\leq C \int_{\R}\chi^2(\sigma)d\sigma.
\end{equation}
So $\|f\|_{\ch^2}\|g_N\|_{L^2}\leq C$.
Remark that, since $f(x)V(x)g_N(x)\geq0$ for all $x\in\R^\nu$, we have for $N$ large enough
\begin{eqnarray*}
\int_{\R^\nu}f(x)V(x)g(x)dx&=&\int_{\R^\nu}\jap{x}^{-(\nu+1)/2}\ln\left(\frac{3\pi}{4}+2N\pi\right)^{(1-\nu)/2}\exp(5|x|/4)\\
& &\chi\biggl(\exp(|x|)-2N\pi\biggr)(1-\kappa(|x|))\sin(\exp(|x|))dx\\
&\geq&C\ln\left(\frac{3\pi}{4}+2N\pi\right)^{(1-\nu)/2}\int_{\R^\nu}\jap{x}^{-(\nu+1)/2}\exp(5|x|/4)\\
& &\chi\biggl(\exp(|x|)-2N\pi\biggr)dx\\
&\geq& C\ln\left(\frac{3\pi}{4}+2N\pi\right)^{(1-\nu)/2}\\
& &\int_{\R}\jap{r}^{-(\nu+1)/2}\exp(5r/4)\chi(e^r-2N\pi)r^{\nu-1}dr\\
&\geq& C\int_{\R}\exp(r/8)e^r\chi(e^r-2N\pi)dr\\
&\geq& C\int_{\pi/4}^{3\pi/4}(\sigma+2N\pi)^{1/8}\chi(\sigma)d\sigma\\
&\geq&C(\frac{\pi}{4}+2N\pi)^{1/8}\int_{\pi/4}^{3\pi/4}\chi(\sigma)d\sigma.
\end{eqnarray*}
So $\lim\limits_{N\rightarrow+\infty}\int_{\R^\nu}f(x)V(x)g_N(x)dx=+\infty$. So $V$ is not $\Delta$-bounded.

\smallskip

\item By a simple calculus, we have 
\begin{eqnarray*}
& &x\cdot\nabla V(x)\\
&=&-\kappa'(|x|)|x|\exp(3|x|/4)\sin(\exp(|x|))\\
& &+(1-\kappa(|x|))|x|\exp(7|x|/4)\cos(\exp(|x|))\\
& &+(1-\kappa(|x|))\frac{3}{4}|x|\exp(3|x|/4)\sin(\exp(|x|)).
\end{eqnarray*}
Let $\chi\in C^1_c(\R,\R)$ such that $\supp(\chi)\subset [0,\frac{\pi}{4}]$, $\chi\geq0$ and $\chi(\pi/8)=1$.
Let $N\in\N^*$ and
\[f(x)=\jap{x}^{-(\nu+1)/2} \]
 and 
\[g_N(x)=\ln\left(\frac{\pi}{4}+2N\pi\right)^{(1-\nu)/2}\exp(-|x|/2)\chi\biggl(\exp(|x|)-2N\pi\biggr).\]
As in \eqref{eq: g_N 2}, we can show that $\|f\|_{\ch^1}\|g_N\|_{\ch^1}\leq C$.
Remark that, $f(x)x\cdot\nabla V(x)g_N(x)\geq0$ for all $x\in\R^\nu$. Since $\kappa\in C^\infty_c$, we have for $N\in\N^*$ large enough
 \[\int_{\R^\nu}f(x)\kappa'(|x|)|x|\exp(3|x|/4)\sin(\exp(|x|))g_N(x)dx=0.\]
 Moreover, we have
 \[\int_{\R^\nu}f(x)(1-\kappa(|x|))\frac{3}{4}|x|\exp(3|x|/4)\sin(\exp(|x|))g_N(x)dx\geq0.\]
 Thus, for $N$ large enough,
\begin{eqnarray*}
\int_{\R^\nu}f(x)x\cdot\nabla V(x)g(x)dx&\geq&\int_{\R^\nu}|x|\ln\left(\frac{\pi}{4}+2N\pi\right)^{(1-\nu)/2}\chi\biggl(\exp(|x|)-2N\pi\biggr)\\
& &(1-\kappa(|x|))\jap{x}^{-(\nu+1)/2}\exp(5|x|/4)\cos(\exp(|x|))dx\\
&\geq&C\ln\left(\frac{\pi}{4}+2N\pi\right)^{(1-\nu)/2}\int_{\R^\nu}\jap{x}^{-(\nu+1)/2}\\
& &\chi\biggl(\exp(|x|)-2N\pi\biggr)|x|\exp(5|x|/4)dx\\
&\geq&C\ln\left(\frac{\pi}{4}+2N\pi\right)^{(1-\nu)/2}\int_{\R^\nu}\jap{r}^{-(\nu+1)/2}\\
& &\chi\biggl(e^r-2N\pi\biggr)r^\nu\exp(5r/4)dx\\
&\geq&C\int_{\R^\nu}\chi\biggl(e^r-2N\pi\biggr)\exp(5r/4)dx\\
&\geq& C\int_{0}^{\pi/4}(\sigma+2N\pi)^{1/4}\chi(\sigma)d\sigma\\
&\geq&C(2N\pi)^{1/4}\int_{0}^{\pi/4}\chi(\sigma)d\sigma.
\end{eqnarray*}
Thus $\lim\limits_{N\rightarrow+\infty}\int_{\R^\nu}f(x)x\cdot\nabla V(x)g_N(x)dx=+\infty$.\\
Therefore $V\notin C^1(A_D,\ch^1,\ch^{-1})$. \qed
\end{enumerate}
Remark that, if $f(x)=\jap{x}^{-(\nu+1)/2}$, $f\in\ch^k$ for all $k\in\N$. This yields that, by the same proof, we can show that $V:\ch^k\rightarrow L^2$ is not bounded for all $k\in\N$.

\subsection{A short range potential in a weak sense}

Now, we will show an example of potential with no decay at infinity for which Theorem \ref{th:756} applies.

We have the following:

\begin{lemma}\label{l:H-1}
Suppose that $V:\ch^1\rightarrow\ch^{-1}$ is a symmetric bounded operator. There exists $u\in\cu$ such that:
\begin{enumerate}
\item if $x\mapsto |x|V(x)$ is in $\ch^{-1}$ then $V\in C^1(A_u,\ch^1,\ch^{-1})$.

\smallskip

\item if there is $\mu>0$ such that $x\mapsto \langle x\rangle^{1+\mu} V(x)$ is in $\ch^{-1}$ then \\
$V\in C^{1,1}(A_u,\ch^1,\ch^{-1})$.
\end{enumerate} 
\end{lemma}
For this type of potential, we can take $u$ of the form $u(x)=x\langle x\rangle^{-\nu-1}$. Note that this $u$ is in $ L^2(\R^\nu)$.

We have the following
\begin{corollary}
Let $V:\ch^1\rightarrow\ch^{-1}$ is a symmetric compact operator. Suppose that there is $\mu>0$ such that $x\mapsto \langle x\rangle^{1+\mu} V(x)$ is in $\ch^{-1}$. Then Theorem \ref{th:756} applies on $(0,+\infty)$.
\end{corollary}

We will give an example of a potential which satisfies assumption of the previous corollary and for which we cannot apply the Mourre Theorem with $A_D$ as conjugate operator.
\begin{lemma}\label{l: example V}
Let $\nu\geq3$ and let $\chi:\R\rightarrow\R$ such that $\chi\in C^3$, $\chi(|x|)=0$ if $|x|>1$ and $\chi'(0)=\chi''(0)=1$. Let 
\[V(x)=\sum\limits_{n=2}^{+\infty}n^{(3\nu-1)/2}\chi'(n^{3\nu/2}(|x|-n)) \quad \text{with a finite sum for each }x.\]
Then
\begin{enumerate}
\item $V:\ch^1\rightarrow\ch^{-1}$ is compact;

\smallskip

\item there is $u\in\cu$ such that  $V$ is of class $C^{1,1}(A_u,\ch^1,\ch^{-1})$;

\smallskip

\item $V$ is not $\Delta$-bounded;

\smallskip

\item $V$ is not of class $C^1(A_D,\ch^1,\ch^{-1})$.
\end{enumerate}
\end{lemma}

\proof[Lemma \ref{l:H-1}]
Let $u(x)=x\langle x\rangle^{-\nu-1}$.
\begin{enumerate}
\item Suppose that $x\mapsto |x|V(x)$ is in $\ch^{-1}$. By \eqref{eq:commutateur V,A}, we can see that, if \\$[qV,u(p)]:\ch^1\rightarrow\ch^{-1}$ is bounded, then $V\in C^1(A_u,\ch^1,\ch^{-1})$.\\
Let $f,g\in C^\infty_c$.
By \eqref{eq:norme H1} and Young inequality, there is $C>0$ such that
\begin{eqnarray}\label{eq:qV H-1}
|(u(p)f,qVg)|&\leq&C\int_{\R^{2\nu}} |u(\xi)|\langle \xi\rangle|\hat{f}(\xi)|\langle \xi-\eta\rangle^{-1}|\widehat{qV}(\xi-\eta)|\langle \eta\rangle|\hat{g}(\eta)|d\xi d\eta\nonumber\\
&\leq&C\|u(q)\langle q\rangle \hat{f}\|_{1}\|(\langle q\rangle^{-1}|\widehat{qV}|)*(\langle q\rangle|\hat{g}|)\|_{\infty}\nonumber\\
&\leq&C\|u\|_2\|\langle q\rangle \hat{f}\|_2\|\langle q\rangle^{-1}\widehat{qV}\|_2\|\langle q\rangle\hat{g}\|_2\nonumber\\
&\leq&C\|u\|_2\|f\|_{\ch^1}\|qV\|_{\ch^{-1}}\|g\|_{\ch^1}.
\end{eqnarray}

We have a similar inequality for $|(qVf,u(p)g)|$. By density, we have the same inequality for all $f,g\in\ch^1$.
Thus $[qV,u(p)]:\ch^1\rightarrow\ch^{-1}$ is bounded which implies that $V\in C^1(A_u,\ch^1,\ch^{-1})$.

\smallskip

\item Suppose that there is $\mu>0$ such that $x\mapsto \langle x\rangle^{1+\mu} V(x)$ is in $\ch^{-1}$. 
In particular, $x\mapsto|x|V(x)$ is in $\ch^{-1}$. Therefore $V\in C^1(A_u,\ch^1,\ch^{-1})$.
As we saw previously, we can deduce that $[\langle q\rangle^\mu qV,u(p)]:\ch^1\rightarrow\ch^{-1}$ is bounded.\\
By a simple calculus, we have:
\begin{eqnarray*}
[qV,u(p)]&=&[\langle q\rangle^{-\mu}\langle q\rangle^\mu qV,u(p)]\\
&=&\langle q\rangle^{-\mu}[\langle q\rangle^\mu qV,u(p)]+[\langle q\rangle^{-\mu},u(p)]\langle q\rangle^\mu qV
\end{eqnarray*}
By the pseudo-differential calculus, we can prove that $\langle q\rangle^{\mu}[\langle q\rangle^{-\mu},u(p)]\langle p\rangle^{1+\nu}$ is a bounded operator. From that, we deduce
\[[\langle q\rangle^{-\mu},u(p)]\langle q\rangle^\mu qV=\langle q\rangle^{-\mu}\langle q\rangle^{\mu}[\langle q\rangle^{-\mu},u(p)]\langle p\rangle^{1+\nu}\langle p\rangle^{-1-\nu}\langle q\rangle^\mu qV.\]
As in \eqref{eq:qV H-1}, since $x\mapsto \langle x\rangle^{-1-\nu}$ is in $L^2(\R^\nu)$, $\langle p\rangle^{-1-\nu}\langle q\rangle^\mu qV:\ch^1\rightarrow\ch^{-1}$ is bounded and $\langle q\rangle^{\mu}[\langle q\rangle^{-\mu},u(p)]\langle p\rangle^{1+\nu}\langle p\rangle^{-1-\nu}\langle q\rangle^\mu qV:\ch^1\rightarrow\ch^{-1}$ is bounded.\\
Thus we can write $[qV,u(p)]=\langle q\rangle^{-\mu}B$ where $B:\ch^1\rightarrow\ch^{-1}$ is bounded. Thus, for all  $\xi$ a real function of class $C^\infty(\R^\nu)$ such that $\xi(x)=0$ if
$|x|<1$ and $\xi(x)=1$ if $|x|>2$, we have 
\begin{eqnarray*}
\|\xi(q/r)[qV,u(p)]\|_{\cb(\ch^1,\ch^{-1})}&=&\|\xi(q/r)\langle q\rangle^{-\mu}B\|_{\cb(\ch^1,\ch^{-1})}\\
&\leq& \langle r\rangle^{-\mu}\|B\|_{\cb(\ch^1,\ch^{-1})}.
\end{eqnarray*}
In particular, by \eqref{eq:qV H-1}, $V$ satisfies \eqref{eq:lr} and, by Proposition \ref{pr:lr}, $V$ is of class $C^{1,1}(A_u,\ch^1,\ch^{-1})$.\qed
\end{enumerate}


\proof[Lemma \ref{l: example V}]
 Remark that we can write $V(x)=\frac{x}{|x|}\nabla W(x)$ where 
 \[W(x)=\sum\limits_{n=2}^{+\infty}\sqrt{n}^{-1}\chi(\sqrt{n}^{3\nu}(|x|-n)).\]
 \begin{enumerate}
 \item Since $\lim\limits_{|x|\rightarrow+\infty}W(x)=0$, by writing $\nabla W(q)=i[p,W(q)]$, $V:\ch^1\rightarrow\ch^{-1}$ is compact.
 
 \smallskip
 
\item  Let $\mu>0$.
 \[
 \int_{\R^\nu}\Bigl||x|^{1+\mu}W(x)\Bigr|^2dx=\sum\limits_{n=2}^\infty n^{-1}\int_{\R^\nu}|x|^{2+2\mu}\chi^2(n^{3\nu/2}(|x|-n))dx
 \]
 Since $\chi(|x|)=0$ if $|x|>1$, there is $C>0$ such that
 \begin{eqnarray*}
  \int_{\R^\nu}\Bigl||x|^{1+\mu}W(x)\Bigr|^2dx&\leq&C\sum\limits_{n=2}^\infty n^{-1}(n+n^{-3\nu/2})^{2+2\mu}\int_{n-n^{-3\nu/2}}^{n+n^{-3\nu/2}}r^{\nu-1}dr\\
  &\leq&2C\sum\limits_{n=2}^\infty (n+n^{-3\nu/2})^{1+2\mu+\nu}n^{-3\nu/2-1}.
 \end{eqnarray*}
 In particular, since $\nu\geq 3$, for $\mu>0$ sufficiently small, this sum is finite, and we can conclude that
 $\jap{q}^{1+\mu}W\in L^2$. Therefore 
\[\jap{x}^{1+\mu}V(x)=\jap{x}^{1+\mu}\frac{x}{|x|}\nabla W=\frac{x}{|x|}\nabla\left\{\jap{\cdot}^{1+\mu}W\right\}(x)-(1+\mu)|x|\jap{x}^{\mu-1}W(x).\]
Since $x\mapsto |x|\jap{x}^{\mu-1}W(x)$ is in $L^2$, $x\mapsto \langle x\rangle^{1+\mu} V(x)$ is in $\ch^{-1}$. \\Thus, by Lemma \ref{l:H-1}, $V\in C^{1,1}(A_u,\ch^1,\ch^{-1})$.

\smallskip

\item Let $N\in\N^*$
\[f(x)=\jap{x}^{-(\nu+1)/2}\quad \text{ and } \quad g_N(x)=N^{\nu/4+1/2}\chi'(N^{3\nu/2}(|x|-N)).\]
Remark that $f\in\ch^2$ and 
\begin{eqnarray}\label{eq: g_N}
\int_{\R^\nu}g_N^2(x)dx&=& N^{\nu/2+1}\int_{\R^\nu}\chi'(N^{3\nu/2}(|x|-N))^2dx\nonumber\\
&\leq&C N^{\nu/2+1}\int_{\R^\nu}r^{\nu-1}\chi'(N^{3\nu/2}(r-N))^2dx\nonumber\\
&\leq& C N^{\nu/2+1}(N+1)^{\nu-1}N^{-3\nu/2}\int_{-1}^1\chi'(t)^2dt\nonumber\\
&\leq& C.
\end{eqnarray}
Thus $\|g_N\|_{L^2}\leq C$. To simplify notation, let $I=[N-N^{-3\nu/2},N+N^{-3\nu/2}]$
Then 
\begin{eqnarray*}
\int_{\R^\nu}f(x)V(x)g_N(x)dx&=&N^{\nu/4+1/2}N^{(3\nu-1)/2}\\
& &\quad\int_{\R^\nu}\jap{x}^{-(\nu+1)/2}\chi'(N^{3\nu/2}(|x|-N))^2dx\\
&\geq&N^{\nu/4+1/2}N^{(3\nu-1)/2}\jap{N+N^{-3\nu/2}}^{-(\nu+1)/2}\\
& &\int_{|x|\in I}\chi'(N^{3\nu/2}(|x|-N))^2dx\\
&\geq&C N^{\nu/4+1/2}N^{(3\nu-1)/2}\jap{N+N^{-3\nu/2}}^{-(\nu+1)/2}\\
& &\int_{r\in I}r^{\nu-1}\chi'(N^{3\nu/2}(r-N))^2dr\\
&\geq&C \jap{N+1}^{-(\nu+1)/2}(N-1)^{5\nu/4-1}\\
& &\int_{r\in I}\chi'(N^{3\nu/2}(r-N))^2dr\\
&\geq&C \jap{N+1}^{-(\nu+1)/2}(N-1)^{5\nu/4-1}\int_{r\in[-1,1]}\chi'(r)^2dr
\end{eqnarray*}
Therefore, since $\nu\geq3$, $\lim\limits_{N\rightarrow+\infty}\int_{\R^\nu}f(x)V(x)g_N(x)dx=+\infty$.

Since $\|f\|_{\ch^2}\|g_N\|_{L^2}\leq C$, then $V$ is not $\Delta$-bounded.

\smallskip

\item By a simple calculus, we have
\[x\cdot\nabla V(x)=|x|\sum\limits_{n=2}^{+\infty}n^{3\nu/2}n^{(3\nu-1)/2}\chi''(n^{3\nu/2}(|x|-n)).\]
Let $N\in\N^*$ and
\[f(x)=\jap{x}^{-(\nu+1)/2}\quad \text{ and } \quad g_N(x)=N^{\nu/4+1/2}N^{-3\nu/2}\chi''(N^{3\nu/2}(|x|-N)).\]
Remark that $f\in\ch^1$ and by \eqref{eq: g_N} $\|g_N\|_{\ch^1}\leq C$.
To simplify notation, let $I=[N-N^{-3\nu/2},N+N^{-3\nu/2}]$. We have
\begin{eqnarray*}
\int_{\R^\nu}f(x)x\cdot\nabla V(x)g_N(x)dx&=&N^{\nu/4+1/2}N^{(3\nu-1)/2}\\
& &\int_{\R^\nu}|x|\jap{x}^{-(\nu+1)/2}\chi''(N^{3\nu/2}(|x|-N))^2dx\\
&\geq&N^{\nu/4+1/2}N^{(3\nu-1)/2}\jap{N+N^{-3\nu/2}}^{-(\nu+1)/2}\\
& &\int_{|x|\in I}\chi''(N^{3\nu/2}(|x|-N))^2dx\\
&\geq&C N^{\nu/4+1/2}N^{(3\nu-1)/2}\jap{N+N^{-3\nu/2}}^{-(\nu+1)/2}\\
& &\int_{r\in I}r^{\nu-1}\chi''(N^{3\nu/2}(r-N))^2dr\\
&\geq&C \jap{N+1}^{-(\nu+1)/2}(N-1)^{5\nu/4-1}\\
& &\int_{r\in I}\chi''(N^{3\nu/2}(r-N))^2dr\\
&\geq&C \jap{N+1}^{-(\nu+1)/2}(N-1)^{5\nu/4-1}\\
& &\int_{r\in[-1,1]}\chi''(r)^2dr.
\end{eqnarray*}

Thus, since $\nu\geq3$, $\lim\limits_{N\rightarrow+\infty}\int_{\R^\nu}f(x)x\cdot\nabla V(x)g_N(x)dx=+\infty$ with\\
 $\|f\|_{\ch^1}\|g_N\|_{\ch^1}\leq C$. Thsu $V\notin C^1(A_D,\ch^1,\ch^{-1})$.\qed
\end{enumerate}

\section{Flow}\label{s:flow}

In this section we make a comment concerning the unitary group
generated by the operator $A_u$ which could be useful in checking the
$C^{1,1}(A_u)$ property, subject that however we shall not pursue
further in this note.  For $A_u$ as in \eqref{eq:A} one may give an
explicit description of $\rme^{\rmi\tau A_u}$ in terms of the classical
flow generated by the vector field $u$ as follows (we refer to
Subsection 4.2 in \cite{ABG} for details). For each $x\in X$ denote
$ \phi_\tau(x)$ the solution of the system
\begin{equation}
\left\{ \begin{array}{l}
\frac{d}{d \tau} \phi_\tau (x) = u\left(\phi_\tau (x)\right)\\
\phi_0(x) = x
\end{array} \right. , \label{eq:flow}
\end{equation}
which exists for all real $\tau$ and $\phi_\tau(x)$ is a $C^\infty$
function of $\tau,x$. Then $\phi_\tau:X\to X$ is a $C^\infty$
diffeomorphism and we have
$\phi_\sigma\circ\phi_\tau=\phi_{\sigma+\tau}$. 

Remark that because $\cf A_u\cf^{-1}=\frac{1}{2}(p\cdot u(q)+u(q)\cdot p)$, $A_u$ is essentially self-adjoint (see \cite[Proposition 4.2.3]{ABG}). Morever, because $\cf^{-1}\ch^s_t\cf=\ch^t_s$, for $u\in\cu$, the $C_0$ groups $e^{i\tau A_u}$ leaves invariant the $\ch^s_t$ spaces (see \cite[Proposition 4.2.4]{ABG}).

 Denote $\phi'_\tau(x)$
the derivative of $\phi_\tau$ at the point $x$, so that
$\phi'_\tau(x):X\to X$ is a linear map with 
$J_\tau(x)=\det\phi'_\tau(x)>0$. Then:
\begin{equation}\label{eq:det}
\cf \rme^{\i\tau A_u}\cf^{-1} f= J_\tau^{1/2} f\circ\phi_\tau 
\end{equation}
where $\cf$ is the Fourier transformation. 

For the operator $A_N$ given by \eqref{eq:anak} it suffices to consider
the one dimensional case, because of the factorization properties
mentioned at the beginning of Section \ref{s:lap}. Then $X=\R$ and
$u(k)=\sin(ak)$. For simplicity, and without loss of generality, we
take $a=1$. Then the system \eqref{eq:flow} has an elementary
solution: if $0\leq x \leq \pi$ then
\begin{equation} 
  \phi_\tau( x) =  
  \arccos \left(\frac{\left( 1 - e^{2\tau} \right) + 
      \cos \left( x \right) \left(1+ e^{2\tau} 
      \right)}{\left( 1 + e^{2\tau}\right) + 
      \cos \left( x \right) \left( 1 - e^{2\tau} \right)} \right)
=2\arctan\left(\rme^\tau\tan\left(\frac{x}{2}\right) \right)
\end{equation}
and similarly outside $[0,\pi]$.  Note that if $x = k\pi$ with $k\in
\Z$ then $\phi_\tau(x) = x$.

\medskip

{\bf Acknowledgements.} This article is based on my master's thesis.  I
thank my advisor, Vladimir Georgescu, for drawing my attention to
Nakamura's paper and for helpful discussions. I thank my doctoral supervisor, Thierry Jecko, for comments which helped me.

\bibliographystyle{alpha}
\bibliography{bibliographie}

\end{document}